\documentclass[prb,twocolumn,aps,floatfix]{revtex4-2}
\usepackage{graphicx,amsmath,xcolor}
\usepackage{hyperref}
\usepackage{subcaption}
\usepackage{float}
\usepackage{svg}
\usepackage[normalem]{ulem}
\usepackage{silence}
\usepackage{mathtools}
\usepackage[T1]{fontenc}
\DeclarePairedDelimiterX\innerp[2]{\langle}{\rangle}{#1\delimsize\vert\mathopen{}#2}%
\DeclarePairedDelimiterX\braket[2]{\langle}{\rangle}{#1\delimsize\vert\mathopen{}#2}%
\DeclarePairedDelimiterX\braketOP[3]{\langle}{\rangle}{#1\,\delimsize\vert\,\mathopen{}#2\,\delimsize\vert\,\mathopen{}#3}%
\DeclarePairedDelimiterX\ketbra[2]{\lvert}{\rvert}{#1\delimsize\rangle\!\delimsize\langle#2}%
\DeclarePairedDelimiterX\outerp[2]{\lvert}{\rvert}{#1\delimsize\rangle\!\delimsize\langle#2}%
\DeclarePairedDelimiterX\projector[1]{\lvert}{\rvert}{#1\delimsize\rangle\!\delimsize\langle#1}%
\usepackage{bibunits}

\defaultbibliography{mainbib}
\defaultbibliographystyle{apsrev4-2}

\begin{document}

    \title{Machine Learning the Disorder Landscape of Majorana Nanowires}
    \author{Jacob R. Taylor}
    \author{Jay D. Sau}
    \author{Sankar Das Sarma}
    \affiliation{Condensed Matter Theory Center and Joint Quantum Institute, Department of Physics, University of Maryland, College Park, Maryland 20742-4111 USA}
\begin{abstract}
We develop a practical machine learning approach to determine the disorder landscape of Majorana nanowires by using training of the conductance matrix and inverting the conductance data in order to obtain the disorder details in the system.  The inversion carried out through machine learning using different disorder parametrizations turns out to be unique in the sense that any input tunnel conductance as a function of chemical potential and Zeeman energy can indeed be inverted to provide the correct disorder landscape.  Our work opens up a qualitatively new direction of directly determining the topological invariant and the Majorana wave-function structure corresponding to a transport profile of a device using simulations that quantitatively match the specific conductance profile. In addition, this also opens up the possibility for optimizing Majorana systems by figuring out the (generally unknown) underlying disorder only through the conductance data.  An accurate estimate of the applicable spin-orbit coupling in the system can also be obtained within the same scheme.
\end{abstract}
\maketitle 
\begin{bibunit}
\textit{Introduction.\textemdash }
Hybrid superconductor-semiconductor (SC-SM) nanowire structures are the most extensively studied systems for creating laboratory Majorana zero modes (MZMs), which are localized non-Abelian excitations that can be used for creating a topological quantum computer. \cite{aghaee2023inas,das2023search,sarma2015majorana}.  A recent experiment \cite{aghaee2023inas} has provided extensive tunnel conductance measurements to make the case for the existence of small and fragile topological regimes where MZMs should exist at the wire ends.  Earlier experiments in such nanowires were strongly adversely affected by unintentional disorder in the system with the MZM-like conductance signatures arising from disorder-induced trivial Andreev bound states (and not from topological MZMs).  \cite{zhang2021large,das2023search,ahn2021estimating,woods2021charge,pan2021quantized,sarma2021disorder,pan2021three,pan2020physical}  Although cleaner samples are used in \cite{aghaee2023inas},  the disorder situation even for this latest experiment is not yet completely clarified, but the small fragile gaps reported in \cite{aghaee2023inas} indicate that random disorder is likely still playing a role. \cite{sarma2023density,sarma2023spectral}  Disorder has thus emerged as the single most important physical mechanism suppressing topology, certainly in Majorana nanowires, but likely in most solid state topological platforms.  Although the possible importance of disorder in suppressing MZM physics was pointed out early \cite{motrunich2001griffiths,lobos2012interplay,akhmerov2011quantized,liu2012zero,adagideli2014effects,bagrets2012class,takei2013soft,brouwer2011probability,rieder2013reentrant,pikulin2012zero,sau2013density,sau2012experimental}, we still do not have any direct information about the disorder in actual samples, hindering progress in the field.  Understanding and controlling unintentional (and thus, unknown) random disorder has become by far the most important problem in the search for MZMs in solid state platforms. \cite{das2023search} Progress toward the realization of topological MZMs necessitates an understanding of the underlying disorder leading to cleaner and better samples. [\cite{das2023search,sarma2023density,sarma2023spectral}. This leads to the other key problem in the field of identifying MZMs in a device based on its transport characterization. This is because, as already mentioned, the unknown disorder leads to transport signatures that are often misinterpreted as 
MZMs, which often exist in a very limited part of parameter space in moderately disordered devices~\cite{adagideli2014effects,sarma2023spectral}. Ultimately, this motivates the other key challenge in the field, which is to identify when an SC-SM nanowire device has been tuned to support MZMs.

In the current work, we introduce an intuitively appealing Machine Learning (ML) approach for figuring out and understanding the disorder landscape in Majorana nanowires using the tunnel conductance data, which are the standard measurements for Majorana nanowires carried out in every experiment (and simulated in the corresponding theories). 
The disorder potential together with other parameters can then be used to quantitatively verify the validity of the model, which in turn can be used to determine if the transport signature indicates a topological superconducting device. The idea is deceptively simple:  The tunnel conductance depends crucially on disorder, and therefore, it should be possible to invert the measured conductance to extract the underlying disorder.  Of course, the uniqueness of such an inverse scattering problem is a key question since, in principle, it is possible for different disorder landscapes to give similar conductance data.  We solve the uniqueness problem a posteriori by showing that input disorder producing the conductance agrees with the output disorder obtained from our ML procedure.  Our ML approach is powerful, and can be used to obtain other quantities entering the MZM physics, and we estimate the applicable spin-orbit coupling (a key parameter directly determining the nanowire topological gap) using our theory. As a matter of principle, the approach we use can be extended to improve estimates of other parameters of the model. The transport profile of such a complete model can be used to verify the model and provide the most direct correspondence so far of whether the specific transport profile corresponds to a topological device. We mention that other types of ML approaches have been used for optimizing gate operations in Majorana nanowires \cite{thamm2023machine}, but our work is totally different since we use ML to solve the inverse scattering problem of extracting the disorder by using tunnel conductance itself as our training data. More significantly, our ML approach, by providing the unknown parameters for the system, enables a direct realistic estimate for the topological invariant as a function of system parameters to decisively ascertain where the system is topological with MZMs and where it is trivial with no topological properties.  This eliminates all subjective judgments about the topological nature of the system as ML itself provides an answer provided sufficient data are used for the training.

\textit{Theory.\textemdash } We use the extensively used minimal 1D model for Majorana nanowires where the system Hamiltonian is given by a 1D BdG equation and the minimal input parameters are the effective mass (m), the spin-orbit coupling ($\alpha$), the Lande g-factor (g), the parent SC gap,  the SC-SM coupling, the self-consistent chemical potential $\mu$, the magnetic field B, and the disorder potential $V_\text{dis} (x)$.  We follow Ref. \cite{sarma2023spectral} where the details (which are standard in the literature to study SC-SM Majorana nanowires~\cite{sarma2015majorana,pan2021three}) can be found.  Note that all quantities other than $V_\text{dis}$ and $\alpha$ are taken to be fixed and known although this constraint can be relaxed in future works (at the cost of needing much more data for training since more unknown parameters there are, larger must be the training data set).  We use, following Ref.~\cite{aghaee2023inas}, a 3 micron long wire for all our results and discretize the BdG equation using a lattice size of 10 nm as in Ref. \cite{sarma2023spectral}.  We provide the details for the theory and the model in the Supplementary Information (S-1 in SI).

Among these SC-SM nanowire parameters, the specific spatial dependence of the disorder potential $V_\text{dis}(x)$ is completely uncontrolled and  varies among devices as well as slowly over time in the same device. In addition, some system parameters such as $\alpha$ depend on the inversion symmetry breaking of the final device structure, which in general is unknown (and most likely varies from device to device). Given a $V_\text{dis} (x)$, we can solve the BdG equation and generate the 4-component tunnel conductance matrix using the KWANT scattering matrix approach. \cite{groth2014kwant}  In addition, we use the spin-orbit (SO) coupling $\alpha$ also as an unknown parameter.  We vary the disorder potential and $\alpha$ to generate, using KWANT, training transport data (i.e. the 4-component conductance matrix G as a function of the magnetic field $B$ and the chemical potential $\mu$) set for our ML algorithm. We provide details of how the conductance matrix $G$ is computed as a function of $B,\mu$ and other parameters in S-1 of SI. The ML algorithm then predicts $V_\text{dis}$ and $\alpha$ that can be used using KWANT to reproduce a test transport data set. Ideally, in the test transport data would come from experiments. In our proof of principle
demonstration, our test transport data is generated using KWANT with a random choice of $V_\text{dis}(x)$ and $\alpha$. The real test of our ML success is whether the conductance generated using the output $V_\text{dis}(x)$ and $\alpha$ is close to the test transport data set.

\textit{Method.\textemdash } Our ML model consists of a{convolutional neural
network (CNN)} created by the package Keras \cite{chollet2015keras} which builds upon tensorflow \cite{tensorflow2015-whitepaper}. In our specific {CNN} (Fig. 1) each
input step consists of a measurement operation, 7 parameters
consisting of a row of X (4 parameters), which are the components of the conductance matrix $G_{\alpha\beta=L,R}$ and of K (3
parameters), which are the parameters bias voltage $V_{bias}$, chemical potential $\mu$ and magnetic field $B$. {The conductance measurements are reshaped into a 3D array based on the values of K columns.}
The 5 Fourier components of $V_{dis}(x)$ and $\alpha$ are organized into the output vector $Y$ of the {CNN} in Fig. 1. As described in S-1 of SI, the conductance matrix $X$ is generated for each instance of $K$ and $Y$ using a KWANT simulation. Later we will also discuss data where we use a $10$ Fourier component model for the disorder. In this case $Y$ would be a 10 component vector. {The convolutional neural network that was chosen is based on AlexNet \cite{krizhevsky2012imagenet} to process the conductance plots visually and to detect hidden structures within them. The device consists of sets of 3D convolutional layers followed by a set of 2D convolutional layers, aggregating conductance measurements for different K parameters \(\mu, B, V_{\text{Bias}}\) (in 3D layers) and \(\mu, B\) (in the 2D layers). The CNN is sufficient for proof of principle but can likely be enhanced with more elaborate methods, such as those based on transformers \cite{liu2021swin}, or an encoder-decoder setup \cite{percebois2023reconstructing}, which may scale better.}
(More ML technical details are given in S-II in SI.)

\begin{figure}[H]
     \centering
     \begin{subfigure}[b]{0.9\linewidth}
         \centering
         \includegraphics[width=\textwidth]{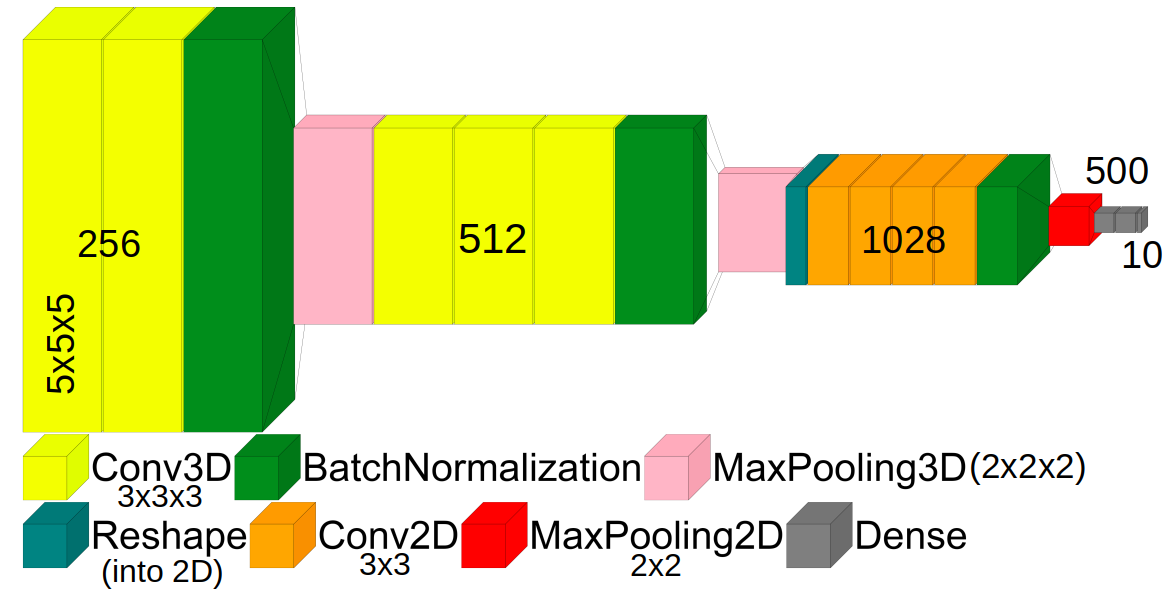}
         \caption{}
         \label{fig:1a}
     \end{subfigure}
     \caption{{(a) Neural network diagram: The model combines a CNN with dense layers. The CNN has 3 convolutional layer sets: two with 256, three with 512, and four with 1028 filters, respectively. Each set is followed by batch normalization and max pooling (2x2x2 for 3D, 2x2 for 2D layers). The first two sets are 3D, starting with a 5x5x5 kernel, then 3x3x3; the third set contains 4 2D layers. A reshaping layer maps to 2D before ending with two 500-size dense layers and an output-sized dense layer. RELU functions add nonlinearity. The output $Y$ includes $\alpha$ only for the 5+1 component model.}}
     \label{fig:NNDiagram}
\end{figure}

\begin{figure}[H]
     \centering
     \begin{subfigure}[b]{0.9\linewidth}
         \centering
         \includegraphics[width=\textwidth]{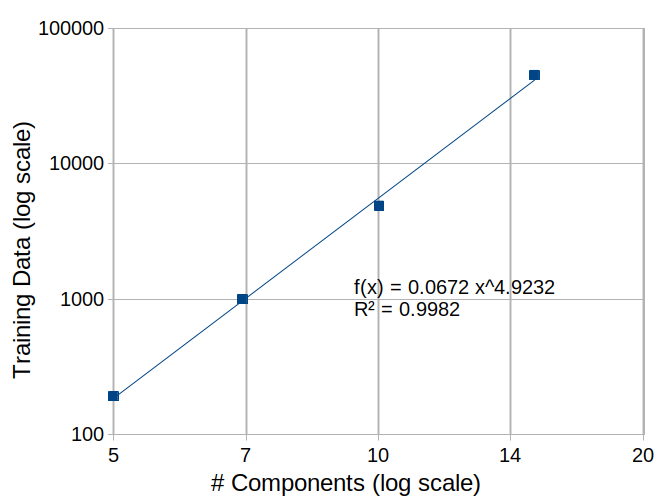}
         \caption{}
     \end{subfigure}
     \caption{{(a) Training (Y) realizations required for $R_2=0.77$ by disorder component count. It's a log-log plot, indicating polynomial scaling with the fit function $f(x)=0.0672x^{4.9232}$ ($R^2=0.9982$), confirming such scaling. A fixed neural network architecture/size is used, shown in Fig. \ref{fig:NNDiagram}, with a constant number of measurements (20 B, 20 $\mu$, and 5 $V_{bias}$ values) except for the 5 components where, to prevent overfitting, 15x15x5 is used. The deviation of the 10 component point is explained by the NN being originally fine-tuned for it, removing this point suggests near-perfect fitting ($R^2>0.99995$). Polynomial scaling indicates potential for n=30, n=50, and n=100. For more on disorder components required for certain fidelities, see S-II-3.}}
     \label{fig:TrainingScaling}
\end{figure}

\begin{figure*}[!htbp]
     \centering
     \begin{subfigure}[b]{0.46\linewidth}
         \centering
         \includegraphics[width=\textwidth]{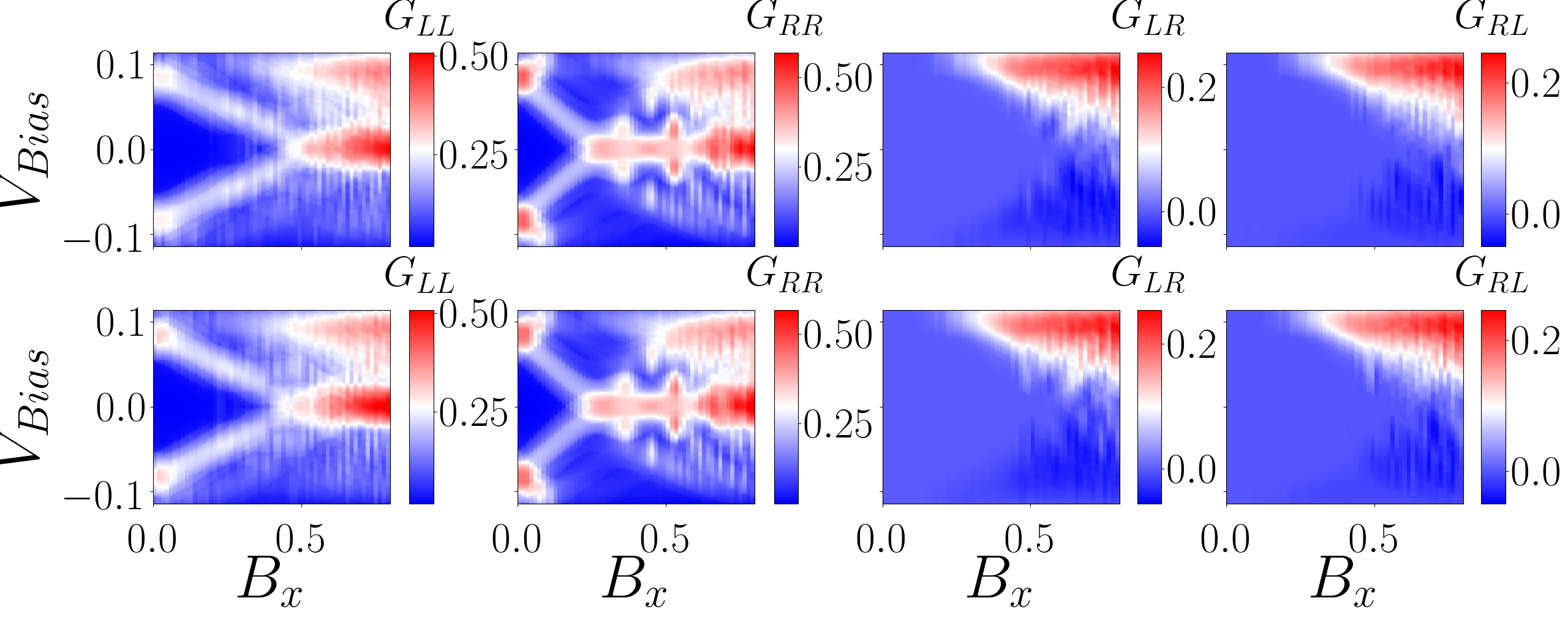}
         \caption{}
         \label{fig:2c}
     \end{subfigure}
          \begin{subfigure}[b]{0.46\linewidth}
         \centering
         \includegraphics[width=\textwidth]{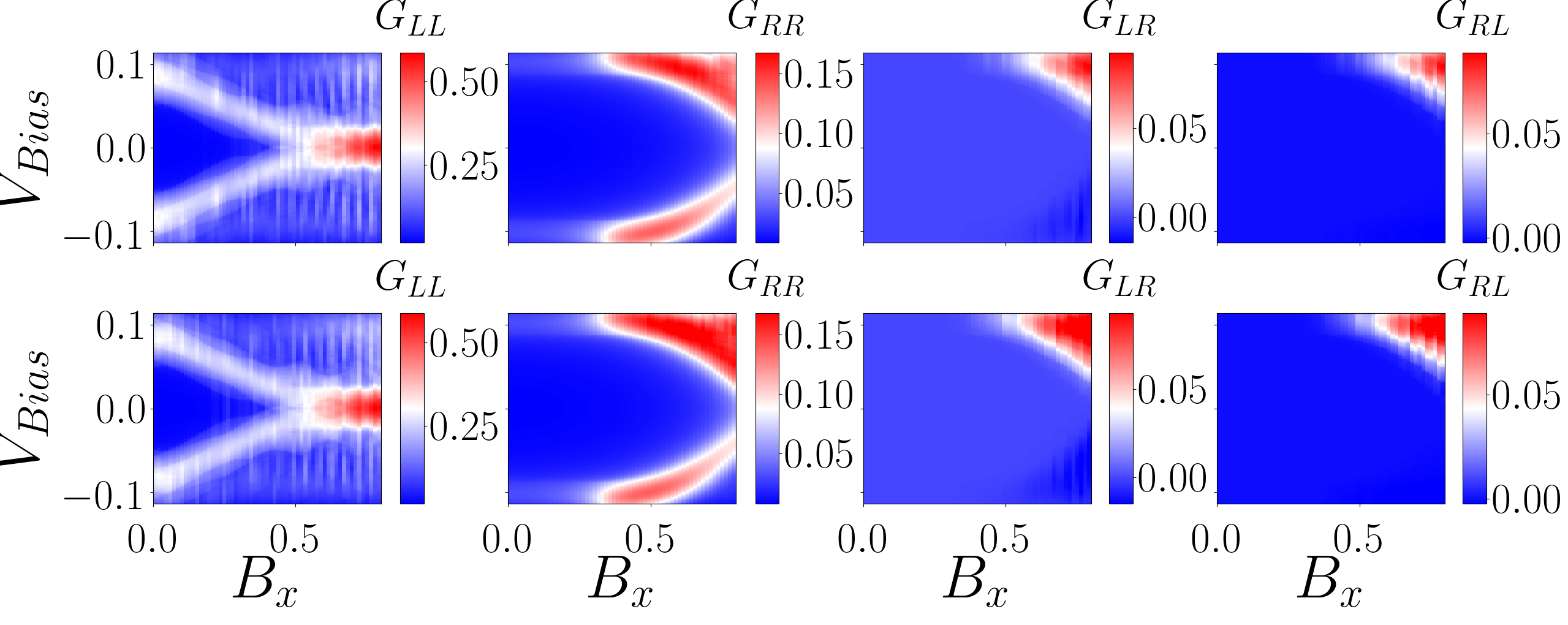}
         \caption{}
         \label{fig:2d}
     \end{subfigure}
          \begin{subfigure}[b]{0.46\linewidth}
         \centering
         \includegraphics[width=\textwidth]{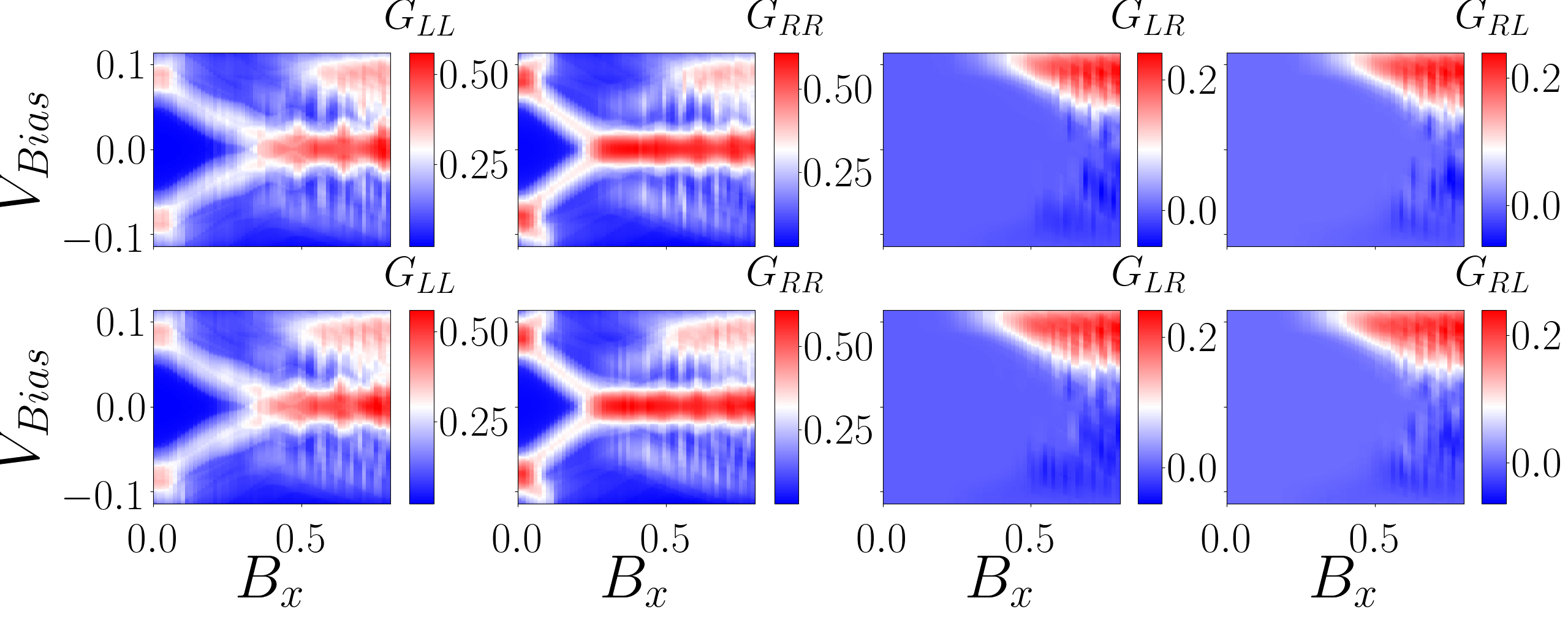}
         \caption{}
         \label{fig:2e}
     \end{subfigure}
          \begin{subfigure}[b]{0.46\linewidth}
         \centering
         \includegraphics[width=\textwidth]{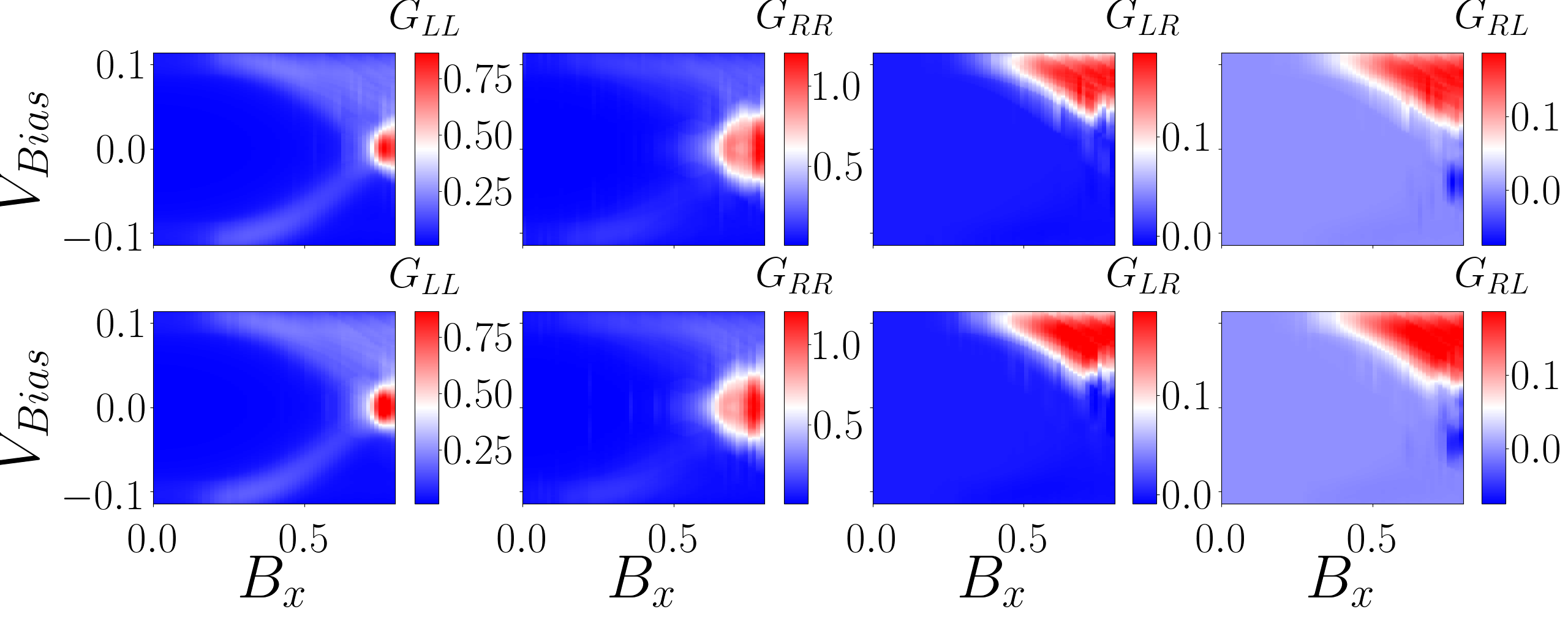}
         \caption{}
         \label{fig:2f}
     \end{subfigure}
     \caption{{Representative input and predictions for 5-component disorder potentials and $\alpha$ in neural network tests. (a-d) two rows: first for expected conductance ($G_{LL}$, $G_{RR}$, $G_{LR}$, $G_{RL}$) from input disorder potentials, and second for conductance measurements from predicted potentials. Each sub-figure represents different disorder and $\alpha$ realizations (Fig. \ref{fig:D1}). The associated topological invariant is in Fig. S7. Inputs and predictions are based on 5-component truncated disorder potentials, with conductance simulated across various $V_{\text{Bias}}$ (meV) and $B_x$ (T) values, at $\mu=0$.}}

     \label{fig:VB1}
\end{figure*}

\begin{figure}[H]
     \centering
      \begin{subfigure}[b]{0.9\linewidth}
         \centering
         \includegraphics[width=\textwidth]{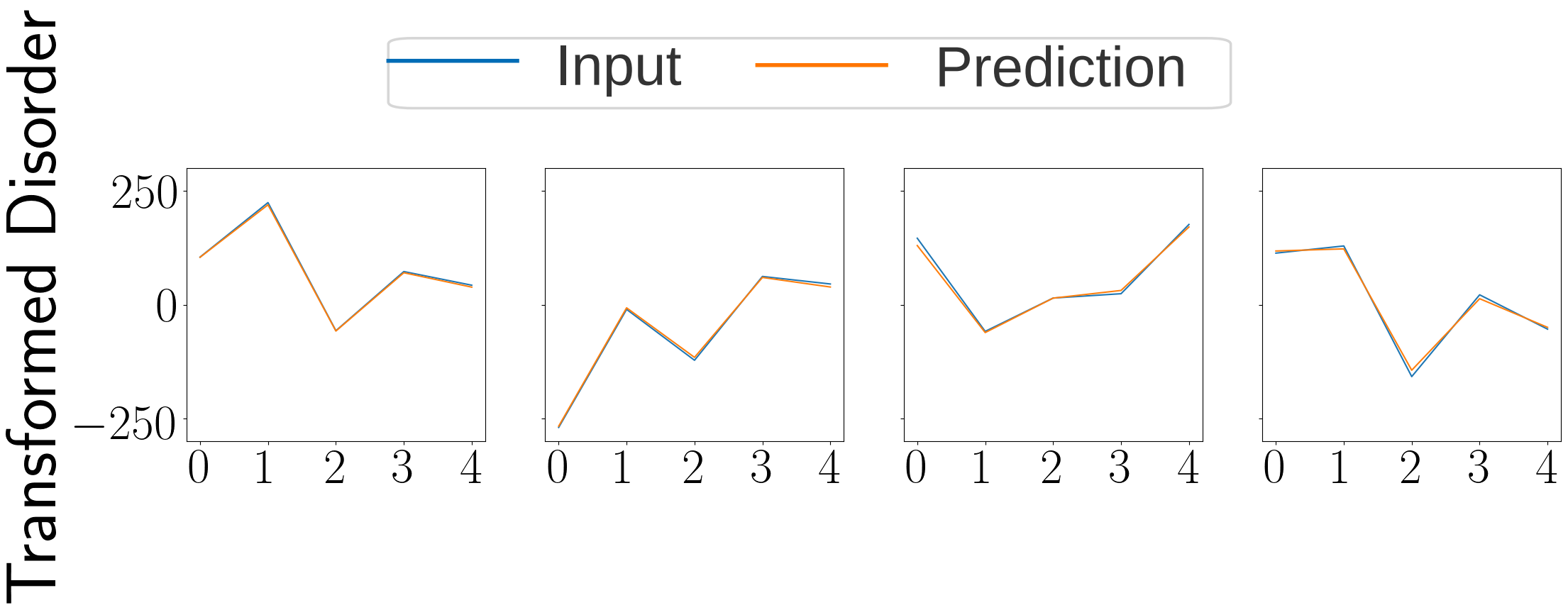}
         \caption{}
         \label{fig:3a}
     \end{subfigure}
      \begin{subfigure}[b]{0.9\linewidth}
         \centering
         \includegraphics[width=\textwidth]{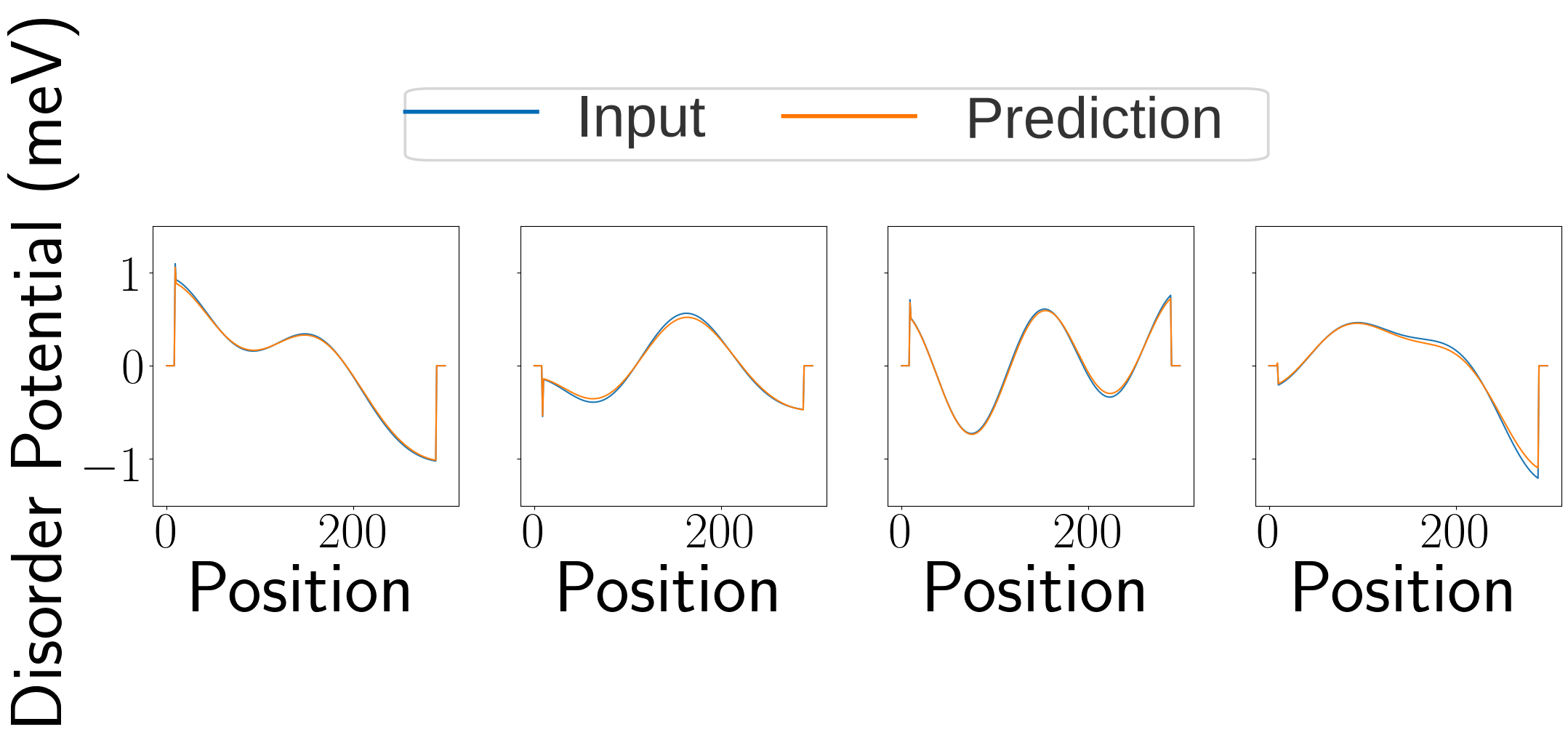}
         \caption{}
         \label{fig:3b}
     \end{subfigure}
     \caption{{Representative ML input and predictions for 5-component disorder potentials. (a) Shows discrete cosine-transformed components for input and predictions. (b) real-space disorder potentials with boundary effects. The corresponding topological invariant is found in {Fig. S7}. Each sample had a disorder amplitude of 1.5meV and correlation length of 3 sites (30nm), with (b) having longer apparent length due to Fourier component truncation.}
     \label{fig:D1}
     }
\end{figure}

\begin{figure}[H]
     \centering
     \begin{subfigure}[b]{0.9\linewidth}
         \centering
         \includegraphics[width=\textwidth]{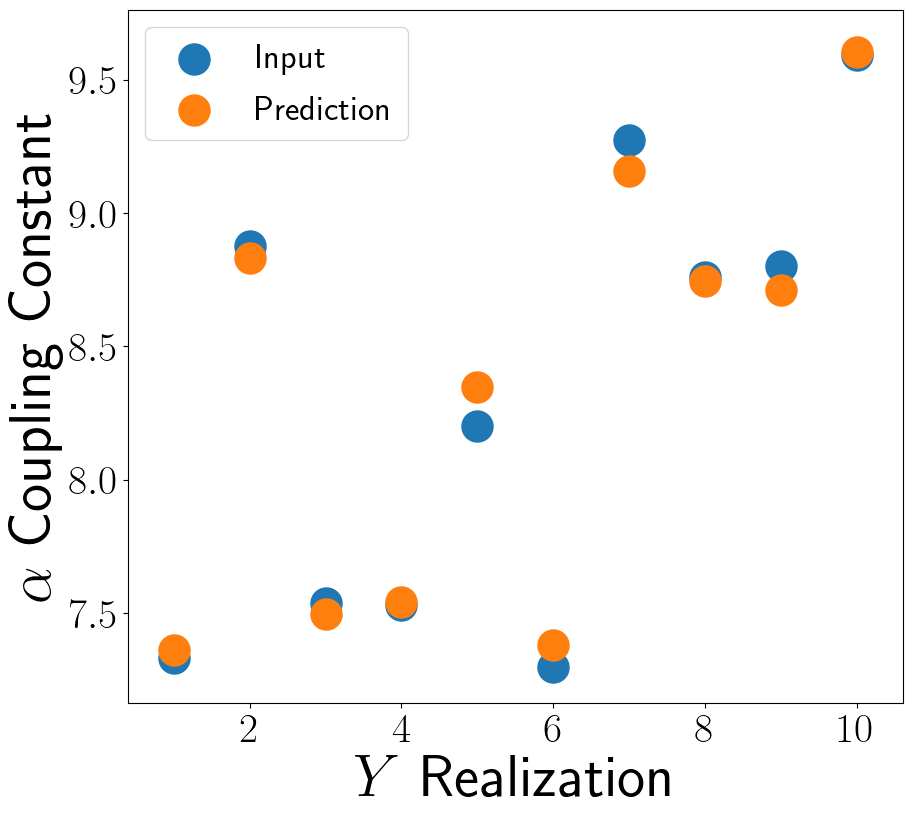}
         \caption{}
     \end{subfigure}
     \caption{Representative samples of $\alpha$ neural network test input and predictions. The blue points representing the input $\alpha$ values (in {meV nm}) and the orange represent the predicted $\alpha$ values.}
     \label{fig:AlphaCompare}
\end{figure}

\textit{Results.\textemdash } We use both  5-component and  10-component representations for our Gaussian random disorder, obtaining very similar results. We measure the {ML} effectiveness using the standard scaled $R_2$ fidelity parameter between the $Y_{test}$ and $Y_{predicted}$ predicted by the trained {CNN}. The standard scaling converts the data used in the training processes into a zero mean, unit variance form. $R_2 = 1$ implies perfect prediction, while $R_2 = 0$ is equivalent to just taking the average output of the testing data, ie., no prediction. This scaling and $R_2$ can be found in details on the package scikit-learn \cite{pedregosa2011scikit}.

To train the model for 5 disorder components we generate a total of 8000
$Y$ realizations for a fixed disorder strength.  We split our data such that 90\% is used in the training process and the remaining 10\% used for testing. Using the 800 test realizations we achieve a standard scaled $R_2$ = {0.991}, which is an excellent fidelity. The standard deviation error in the prediction is {$\Delta V_{\text{dis}}(x_i)=\pm 0.0350$ meV and $\Delta \alpha=\pm 0.11$ \text{meV} \text{nm}.}  {For this model we find it sufficient to use only 5 bias values of conductance data (i.e. only use 5 $V_{bias}$ values in $K$) to obtain the $0.991$ fidelity already.} We emphasize that the ML algorithm predicts $V_\text{dis}$, $\alpha$ accurately enough that {the} predicted values can be used in KWANT to match a test conductance quite accurately as seen in Fig.~\ref{fig:VB1}. The good $R_2$ value of the predictions of the {CNN} are apparent from comparing the test and predicted Fourier components of disorder and SO coupling seen in Figs.~\ref{fig:D1} and ~\ref{fig:AlphaCompare}.
(More results are given in the S-III in the SI.) The accuracy of the predicted parameters implies that the $Y$ values predicted by the {CNN} can also be used to generate the topological invariant in parameter (i.e. $K$) space. To test this we compared the topological visibility in $B-\mu$ space between 
a test case and predicted case and found good agreement as seen in Figs. {S7 S8 S9} in the SI. {Further, we find that our model is extremely resistant to experimental errors in conductance measurements to the point where a Gaussian error of magnitude 0.01 $e^2/h$ yields a $R_2=0.99$ and 0.05 $e^2/h$ yields $R_2=0.98$, even going to 0.1 $e^2/h$ only decreases to $R_2=0.97$. In context the noise floor in experiment\cite{aghaee2023inas} is about 0.001 $e^2/h$ making conductance errors effectively irrelevant.}

In {Fig. S4}, we show the results of our 10-component disorder ML results. To train for 10 disorder components we generate a total of {50000} disorder realizations, 90\% of which is used in the training process and the remaining 10\% used to test. We use a fixed {$\alpha = 8.0 \text{meV nm}$}. Using the {5000} test realizations we achieve a standard scaled {$R_2 = 0.95$}, which is {great}, but can be improved with more data sets. The standard deviation error in the prediction is $\Delta V_{\text{dis}}(x_i)=\pm 0.117$ meV. Due to computational limitations we only use {5 $V_\text{Bias}$ values} (using more values of $V_\text{Bias}$ should increase $R_2$). This lower fidelity is likely due to the need for more training {and potentially a slightly} larger $K$ matrix since differentiating among 10 component potentials is a more computationally challenging task. Representative results are shown in {Fig. S4}, and again, visual inspection shows excellent agreement between the input/output data, verifying the success of our ML approach.
When considering further scalability, a couple of factors warrant attention. {First, we show that the data requirements do not increase exponentially} making it feasible to compute additional components. Furthermore, since we anticipate that the low-frequency disorder components will be dominant, the number of extra components needed to accurately represent the device's physics may be limited. {We do an extended assessment of the number of components required within S-II.3. }

 Additionally, our method's disorder prediction can be verified through the use of additional conductance measurements. While we do not expect additional effects to be necessary, in principle, if our model proves inadequate for describing the physics of an experimental device, it is possible that our learning scheme could produce an invalid disorder prediction. This possibility can be significantly mitigated by conducting measurements with a new $\textbf{K}{\text{full}}'$ (see S-II), which is not integrated into the ML scheme, and comparing the results to the theoretical conductance values generated by KWANT for the respective disorder prediction. This allows for the experimental validation of the machine-learned system parameters. Subsequently, one could iteratively introduce additional potential physical effects into the model until it successfully passes all $\textbf{K}_{\text{full}}'$ tests, leading to a better understanding of the device's underlying physics.

\textit{Conclusion.\textemdash }We introduce a machine learning approach to extract unknown system parameters, particularly the disorder landscape, from the simulated (or measured) tunnel conductance data in hybrid SC-SM Majorana nanowire structures.  We validate the approach by using simulated conductance data as the training set, establishing that such training leads to strongly predictive results for both disorder and spin-orbit coupling using as few as 3 parameters (i.e. $K$) for the  conductance data used in our simulations. The accuracy of the predicted results can be verified by comparing the predicted conductance as well as  topological visibility profiles to the test profiles. The predicted profiles, which are functions of multiple parameters such as $V_{bias}$ and $\mu$ are generated by KWANT from the parameters that are predicted by ML to match a given test conductance data. The topological visibility profile as a function of 
 $B$ and $\mu$ is also generated by KWANT for test conductance data generated by KWANT.  Using experimental conductance data (and a bigger computer), one should be able to generalize our approach to obtain all the relevant parameters for Majorana nanowires, not only the disorder and the SO coupling as we do, but also the g-factor, the SC gap, the number of occupied subbands, etc. since our ML approach is general, and only requires as inputs sufficient amount of conductance training data which are easy to obtain both experimentally and theoretically.
 
 Our method, given enough training data and computing resources,  should in principle be able to decisively indicate, just through our ML protocol,  whether a set of conductance data in a particular sample indicates an underlying topological (or trivial) system with MZMs (or not). In particular, by determining all the relevant unknowns, one can simulate a device and calculate its topological invariant directly using the output parameters, decisively indicating whether the sample is or is not topological. 

\section{Acknowledgement}
We thank Maissam Barkeshli for teaching a machine learning course which helped us in developing our approach presented in this paper.  This work is supported by the Laboratory for Physical Sciences.

\putbib
\end{bibunit}
\begin{bibunit}
\clearpage
\pagebreak
\widetext
\begin{center}
\textbf{\large Supplemental Materials: Machine Learning the Disorder Landscape of Majorana Nanowires}
\end{center}
\setcounter{equation}{0}
\setcounter{figure}{0}
\setcounter{table}{0}
\setcounter{page}{1}
\makeatletter
\makeatother
\renewcommand{\theequation}{S\arabic{equation}}
\renewcommand{\thefigure}{S\arabic{figure}}
\renewcommand{\bibnumfmt}[1]{[S#1]}
\renewcommand{\citenumfont}[1]{S#1}
\setcounter{section}{0}
\renewcommand{\thesection}{S-\Roman{section}}
\renewcommand{\thesubsection}{\thesection.\arabic{subsection}}
\section{Details of theory}
\begin{figure*}[h]
     \centering
     \begin{subfigure}[b]{0.45\linewidth}
         \centering
         \includegraphics[width=\textwidth]{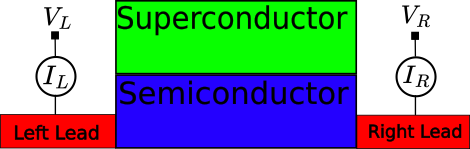}
         \caption{}
         \label{fig:S1b}
     \end{subfigure}
     \caption{Sketch of Majorana nanowire device. The device consists of a NS junction with a layer of superconductor on top of a layer of semiconductor, with opposite ends of the semiconductor connected to left and right leads respectively \cite{pan2020physical}. The differential conductances, which are measured, are defined as $G_{\alpha \beta}= dI_\alpha/dV_\beta$ for $\alpha =\beta$ and $G_{\alpha \beta}= -dI_\alpha/dV_\beta$ for $\alpha \neq \beta$. Local conductances $G_{LL}$ and $G_{RR}$ are measured from their respective leads through the junction (to the superconductor), while non-local conductances $G_{RL}$ and $G_{LR}$ are measured from the right lead to the left lead or the reverse respectively.}
     \label{fig:DeviceSketch}
\end{figure*}

We model the 1D semiconductor Majorana nanowire using a Bogoliubov-de Gennes  Hamiltonian \cite{lutchyn2010majorana}:
\begin{equation}
H=\left( -\frac{\hbar^2}{2m^*}\partial_x^2-i\alpha \partial_x \sigma_y - \mu +V_{\text{dis}}(x) \right)\tau_z\\ +\frac{1}{2}g \mu_B B\left(\sigma_x\cos\theta+\sigma_y\sin\theta \right)+\Sigma(\omega)
\label{eqn:1}
\end{equation}
where $$\Sigma(\omega)=-\gamma \frac{\omega+\Delta_0\tau_x}{\sqrt{\Delta_0^2-\omega^2}}$$ is the self-energy generated by integrating out the superconductor that is proximity coupled to the semiconductor~\cite{sau2010robustness}. The above Hamiltonian is a $4\times4$ matrix comprised of $\sigma_{x,y,z}$, which are $2\times2$ Pauli matrices associated 
with the spin degree of freedom, and $\tau_{x,y,z}$, which are $2\times2$ Pauli matrices associated with the superconducting particle-hole degree of freedom. The above Hamiltonian is written in a basis where the Bogoliubov quasiparticles of the superconductor are described by a Nambu spinor with a structure $\psi(x)=(u_{\uparrow}(x), u_{\downarrow}(x),-v_{\downarrow}(x),v_{\uparrow}(x))^T$, where $\uparrow,\downarrow$ refer to spin and $u,v$ refer to particle-hole components of the quasiparticle. The frequency $\omega$ in the self-energy is equivalent to the energy (in units where $\hbar=1$) of the Bogoliubov quasiparticle in consideration. While, this requires a self-consistent solution for eigenstates in a closed system, the $\omega$ dependence of $\Sigma$ does not present any additional complication in the solution of transport problems in open systems using KWANT~\cite{groth2014kwant}. Specifically, KWANT solves the transport problem of the Majorana nanowire by computing the scattering matrix of the system in the particle-hole space between the two leads shown in Fig.~\ref{fig:DeviceSketch}. 
The scattering matrix can be used to compute the conductance matrix $G_{\alpha\beta}$ using the well-known Blonders-Tinkham-Klapwijk relations~\cite{groth2014kwant}. In addition, one can use the reflection matrix from either of the lead ends to compute the topological visibility $T_{IV}$~\cite{akhmerov2011quantized,pan2021three}.

The actual semiconductor device is a complicated three dimensional semiconductor system. Despite this, the transport properties computed by KWANT \cite{groth2014kwant} using the model Hamiltonian Eq.~\ref{eqn:1} captures most qualitative features of the transport properties of the device over the relevant ranges of gate and bias voltages provided the parameters $\alpha$, $\mu$ etc are taken to be fitting parameters. Therefore, the parameters $m^*$, $\alpha$, $\ldots$ should ideally be viewed as fitting parameters to transport properties of devices. In practice~\cite{pan2021three,woods2021charge,sarma2023spectral}, the parameters such as effective mass $m^*=0.03 m_e$ and a nominal (details later) value of the spin-orbit coupling $\alpha=8.0\,meV-nm$ are chosen to fit non-superconducting transport characterization of the InAs nanowire device. The superconducting pairing potential coupling parameter $\gamma=0.15\,meV$, the Lande g-factor $g=25$ are chosen to qualitatively reproduce transport properties in the superconducting phase. The intrinsic superconducting pair potential $\Delta_0=0.12\,meV$ is chosen according to be the superconducting gap of the Al superconductor. In addition, to model transport in the system given by the Hamiltonian in Eq.~\ref{eqn:1}, we choose a temperature $T=50mK$, and device length $L=3\,\mu m$. 
The finite temperature is implemented in the system by convolving the zero temperature conductance as a function of $\omega$ with the derivative of a Fermi function at temperature $T$. For this purpose, we generate the zero-temperature conductance using KWANT at an $\omega$ grid with 151 points between $\pm 0.04$ meV. 
Finally, for the purpose of numerical simulations we discretize the Hamiltonian in Eq.~\ref{eqn:1} with a lattice scale of a=10 nm. These parameters are chosen from recent work \cite{sarma2023spectral}, which was focused on understanding the recent experiments from the Microsoft group \cite{aghaee2023inas}. 

 The disorder potential generated by impurities in the device is also included in the Hamiltonian $H$ through the function 
 $V_{\text{dis}}(x)$. The spatial disorder potential is assumed to  randomly varying with position $x$ across the length of the wire with a Gaussian distribution with an amplitude $1.5\,\text{meV}$ and correlation length of $30\,nm$. 
 For the purposes of simplifying the ML complexity by reducing the number of parameters, we eliminate (i.e. set to 0) all but $N_{dis}$ of the Fourier components of the disorder potential. For our calculations we will use $N_{dis}=5$ or $10$.
 In addition,
we also include barrier potentials $V_{\text{Barrier}}^L$ and $V_{\text{Barrier}}^R$ as $\delta$ function contributions at the ends of the wire to $V_{\text{dis}}(x)$.

Out of the parameters that go into $H$, the spacial profile of the disorder potential $V_\text{dis}(x)$, which  is specified in our model by the $N_{dis}$ Fourier components, is completely uncontrolled and varies between devices as well as slowly over time. In addition, some parameters such as $\alpha$ depend on the inversion symmetry breaking of the final device structure, where it is difficult to interpret the transport data.
In this work, we will vary the disorder potential and spin-orbit coupling $\alpha$  (around the aforementioned nominal value) to generate, using KWANT, training transport data set as a function of  chemical potential $\mu$, magnetic field  $B$ and bias voltage $V_{bias}$ (related to $\omega$ in $H$) to be fit by an ML algorithm. To keep the computation in this work simple, we assume the other parameters in the Hamiltonian $H$ are fixed at values estimated from experiment. The ML algorithm will then predict the Fourier components of $V_\text{dis}$ and spin-orbit coupling $\alpha$ from transport  that can be used using KWANT to match a test transport data. Ideally, the test transport data would come from experiment though that would likely require more fit parameters. In our proof of principle demonstration, our test transport data is generated using KWANT using a random choice of disorder configuration and $\alpha$.


\section{Details of Machine Learning}
\label{S:ML}
\subsection{Training Data}
The training data was generated within KWANT for many different randomly generated $V_\text{dis}(x)$ and $\alpha$ configurations (called Y realizations). For each Y realization we calculate a set of $G_{\text{RR}}$, $G_{\text{LL}}$, $G_{\text{RL}}$, $G_{\text{LR}}$ values for different measurement configurations. A Y realization consists of generating a random $\alpha$ from a uniform distribution between 6.4 and 9.6, along with a random Gaussian disorder potential. {In particular the initial disorder potential was generated by random sampling of a Gauassian distribution of standard deviation 1.5meV at each site and then convolving these samples together with decaying Gaussian function for a fixed correlation length. In our case the correlation length was 3 sites (30nm).} The disorder potential $V_\text{dis}(x)$ refers to disorder within the $\mu$ chemical energy taking $\mu \rightarrow \mu + V_{dis}(x)$.  The disorder potential is transformed with a discrete cosine transform, and only the nth lowest frequency components are kept. This gives us $c_1 ... c_n$, the transformed disorder potential components. The transformation is then undone to get the disorder potential in real space corresponding to these components. {Along with deterministically introducing boundary effects, we subtract a spacially constant factor updating $V_{dis}(x)$ such that the effective 0th component induces a random $\mu$ shift with standard deviation of 0.01meV. This shift can be determined from the disorder components.} {The} truncated, reverse transformed {then modified} $V_{dis}(x)$ is what will be used by KWANT. From the transformed components we define a vector for each Y realization:
$$Y=\begin{bmatrix}c_1 & \ldots & c_n & \alpha \end{bmatrix}$$
This is done for many different Y realizations, each time having the potential along with $\alpha$ numerically simulated to evaluate a sequence of conductance measurements. The measurement configurations were recorded in a matrix $K$ where each row would record a particular setup of experimentally tunable parameters labelled as j:

$$\textbf{K}_{\text{full}}=\begin{bmatrix}\mu^j & B^j & V^{L,j}_{\text{Barrier}} & V^{R,j}_{\text{Barrier}} &\theta^j& V^j_{\text{Bias}}\end{bmatrix}$$
where $B$ is the applied magnetic field, $\mu$ is the chemical potential parameter controlled by the plunger gate voltage, $V^L_{\text{Barrier}}$ and $V^R_{\text{Barrier}}$ are the left and right barrier voltages respectively, $\theta$ is the angle of the B field relative to the direction of the wire, and $V_{\text{Bias}}$ is the bias voltage. All these are parameters that are readily varied in a transport experiment. The Hamiltonian that is described 
in the previous section, which is used in the KWANT simulation contains all of these parameters except $\theta$,  which has been assumed to be zero for simplicity. The possibility of varying all these parameters to study transport provides the potential of generating a huge volume of transport data to characterize a specific device. In our case, we found it either unnecessary or beyond our limited computational resources to use transport data in the entire space to estimate the disorder in the device. As such we fixed $V^{L}_{\text{Barrier}}=V^{R}_{\text{Barrier}}=15$ {mV} and $\theta=0$. The simplified $\mathbf{K}$ we use is as follows:  

$$\textbf{K}=\begin{bmatrix}\mu^j & B^j & V^j_{\text{Bias}}\end{bmatrix}$$

Furthermore, {many $V_{\text{Bias}}$ measurements} turn out to not be necessary to fit the transport data. {We only use  5 values of $V_{\text{Bias}}$ from our transport data to fit and include within our $K$ matrix. For these calculations, $V_{\text{Bias}}$ takes 5 values between $\pm 0.04$ mV depending on the realization index $j$.}
Each row of K when fed into KWANT generates a new set of conductance measurements, we put these conductances into a matrix we call $\textbf{X}$ as follows: 

$$\textbf{X}=\begin{bmatrix} G_{\text{RR}}^j& G_{\text{LL}}^j & G_{\text{RL}}^j & G_{\text{LR}}^j\end{bmatrix}$$

Putting this all together we get our training data generating function as follows: $$F_{\text{Gen}}(Y;\textbf{K})=\textbf{X}$$ 

Note that all the different Y realizations have the same K matrix. It can be seen then that our objective is to find a function that can take in $\textbf{X}$ and $\textbf{K}$, and give us $Y$. Put another way we wish to invert our generator function to find the following: 

$$F^{-1}_{ML}(\textbf{X};K)=Y$$


In terms of the other components, we varied $\mu$ between 0 meV and 0.6 meV (with {20} steps), $B_x$ between 0 T and 0.8 T (with {20} steps). {Except in the case of the 5 disorder component result where we used the same range but with 15 steps instead.} Overall our $\textbf{K}$ matrix had {400}*$N_{V_\text{Bias}}$ rows, as we do not feed all the $V_{\text{bias}}$ data into the neural network. {$N_{V_\text{Bias}}=5$ for 5, 7, 10 and 15} component versions.

In summary we generate many random realizations of $V^i_\text{dis}(x)$ and $\alpha^i$, cosine transform each $V^i_\text{dis}(x)$ and truncate to either 5 or 10 components to form a $Y^i_\text{input}$. Using this we calculate $F_{\text{Gen}}(Y^i_\text{input};K) = X^i_\text{input}$, where within KWANT $Y^i_\text{input}$ is inverse cosine transformed back into real space. The $X_\text{input}^i$, $Y^i_\text{input}$ pairs are split between training $X_\text{train}^i$, $Y^i_\text{train}$ and testing $X_\text{test}^i$, $Y^i_\text{test}$.  Using $X_\text{train}^i$, $Y^i_\text{train}$ we train a neural network to be able to generate a prediction for $Y$ given an $\mathbf{X}$. We assess the validity of our model by comparing $Y^i_\text{test}$ to our neural network prediction $F_{ML}^{-1}(\mathbf{X}_\text{test}^i;K)=Y^i_\text{{Pred}}$. 

\subsection{The Neural Network}

{Our machine learning model consisted of a convolutional neural network (CNN) created the package Keras \cite{chollet2015keras} which is a package that builds upon tensorflow \cite{tensorflow2015-whitepaper}. A diagram of the neural network can be seen in Fig. 1 of the main text. The convolutional neural network takes input formatted into a 3D array of 4 measurements where the indexes refer to measurements with equal $B$, $\mu$ and $V_{Bias}$. The input begins with 4 channels referring to the 4 different measurements performed for each line of the K matrix. This network takes the form of 3 sets of convolutional layers followed by 2 dense layers. The convolution of all 3D convolutional layers is between the 3 different measurement configuration parameters listed above. The first two convolutional layer sets are 3D and over all 3 parameters, while the last set is 2D only convolving over $B$, $\mu$ because at this point we have pooled away the $V_{Bias}$ index. The first set of convolutional layers consists of a 3D convolutional layer of kernel 5x5x5 followed by a 3D convolutional layer with kernel 3x3x3. Both of the layers within the first convolutional layer set have 256 filters. All convolutional layer sets are followed by batch normalization layers and then max pooling layers. The max pooling layers are always 2x2x2 or 2x2 depending on whether they are connected to a 2D or 3D convolutional layers. The second convolutional layer set consists of 3 3D convolutional layers all with kernels 3x3x3 and 512 filters. The final set of convolutional layers is 4 2D convolutional layers with kernel 3x3x3 and 1028 filters. In between the second and third set is a reshaping layer to convert the 3D data into 2D. The network concludes with 2 dense layers with 500 nodes each. The network was optimized using adam's method as our optimizer and mean squared error as our loss function. We performed sets of 5-15 epochs starting at learning rate 1E-4 and then divided by 10 the rate at each additional set of sweeps. 3-4 sets were performed depending on whether the validation fidelity was still increasing or not. The convolutional network was chosen due to it's ability to process the grid of measurements in a vision-like manner where emergent structure may be visible and thus give the network information about the disorder. Originally we had tried a recurrent neural network (RNN), A RNN would allow the addition of more measurements to not decrease the ability of the previous measurements to perform predictions. While this previous network did work achieving $R_2=0.98$ for 5 components and $R_2=0.79$ for 10 components, it became clear the convolutional neural network was able to perform better. It is possible a transformer based similar method to the RNN could potentially work while yielding this benefit.}  


We measure the effectiveness of our model using the standard scaled $R_2$. This is calculated between $\{Y^i_\text{{test}}\}$ and $\{Y^i_\text{{Pred}}\}$. The standard scaling converts the data used in the training processes into a 0 mean, unit variance form. A $R_2=1$ implies perfectly predicting the tests, while $R_2=0$ is equivalent to just taking the average output of the testing data. This scaling and $R_2$ can be found within the package Scikit-learn \cite{pedregosa2011scikit}.

\subsection{Computational Resources and Scaling}
In terms of computational resources, the generation of training data proves to be significantly more demanding than the training process itself. Fortunately, the task of training data generation exhibits an "embarrassingly parallel" nature. Each set of $N_{V_{Bias}}$ rows in matrix K can be computed in parallel with minimal overhead. For our specific setup, generating conductance data for $N_{V_{Bias}}$ rows in K took approximately 12 seconds per set using a single core. With a total of {400} sets of rows and 50,000 different Y realizations, this amounts to a total processing time of {67} thousand core hours. Considering that many supercomputers boast tens of millions of cores, our approach can certainly be scaled up efficiently.{Within the main text in Fig. 2 we show clear polynomial scaling. Using this scaling we expect targetting an $R_2=0.77$ for 30 components would require 1.2 million core hours and 50 would require 15 million core hours. Both of these are well into the realm of achievable, 1.2 million core hours for context is about how many core hours are available on University of Maryland's Zaratan per day. In principle using all of Zaratan we could complete this computation in a single day.

In terms of the amount of components required to achieve certain fidelity in terms of the original disorder, this is dependent on the length of the wire and the disorder length. Defining a inner product fidelity $F=\frac{\vec{V}_{dis}^{exact}\cdot \vec{V}_{dis}^{pred}}{\sqrt{(\vec{V}_{dis}^{exact}\cdot \vec{V}_{dis}^{exact})(\vec{V}_{dis}^{pred}\cdot \vec{V}_{dis}^{pred})}}$, we find to achieve an F=0.95 with our disorder length of 3 sites (30nm) and $3\mu m$ wire length, we need 54 disorder components. This is achievable. However, if the disorder correlation length was longer such as in \cite{aghaee2023inas} the number of components would decrease quite significantly. In this experiment the minimum estimate of their disorder length is more than 7 sites, if we had a correlation length of 7 sites then the number of components to achieve F=0.95 is 29 and to achieve F=0.97 is 33. This implies again that the number of components required is certainly achievable by a wide margin. We provide figures describing the number of disorder components required for different wire lengths and correlation lengths in Fig. \ref{fig:CompReq}}

\begin{figure*}[h]
     \centering
     \begin{subfigure}[b]{0.70\linewidth}
         \centering
         \includegraphics[width=\textwidth]{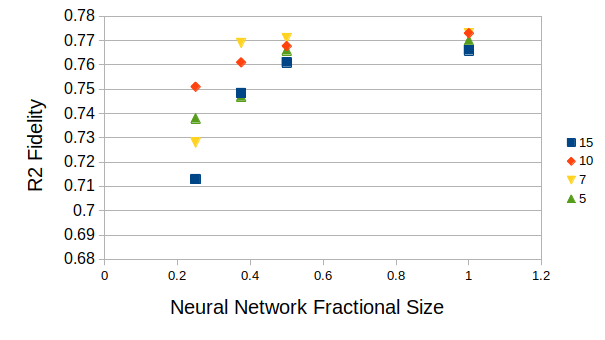}
         \caption{}
     \end{subfigure}
     \caption{{(a) $R_2$ of the neural network for different neural network sizes and number of disorder components. Starting with all the different number of disorder components having a $R_2\approx 0.77$ we show how the $R_2$ decreases as the convolutional layers number of filters shrink. The neural network used is identical to that in Fig. 1 except all the convolutional layers number of filter are multiplied by the fraction on the x-axis. We find halfing the network causes less than 0.01 decrease in fidelity implying we are far into the region of diminishing returns. Further there seems to be no obvious correlation between number of disorder components and required neural network size to achieve a fixed $R_2$. From this we can conclude that the neural network size must scale, if it does, very slowly and certainly not exponentially.}}
     \label{fig:ScalingNN}
\end{figure*}

\begin{figure*}[h]
     \centering
     \begin{subfigure}[b]{0.32\linewidth}
         \centering
         \includegraphics[width=\textwidth]{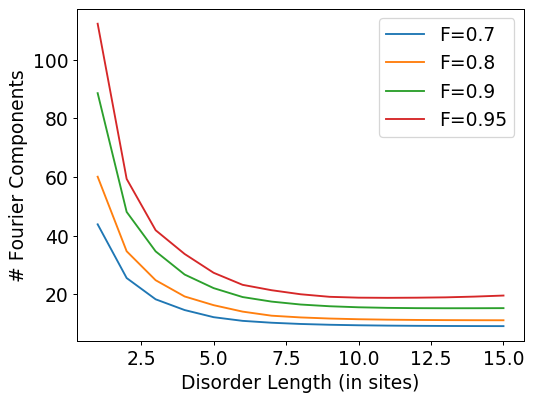}
         \caption{}
     \end{subfigure}
          \begin{subfigure}[b]{0.32\linewidth}
         \centering
         \includegraphics[width=\textwidth]{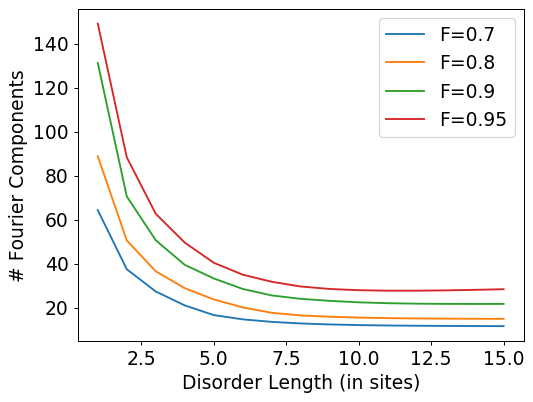}
         \caption{}
     \end{subfigure}
          \begin{subfigure}[b]{0.32\linewidth}
         \centering
         \includegraphics[width=\textwidth]{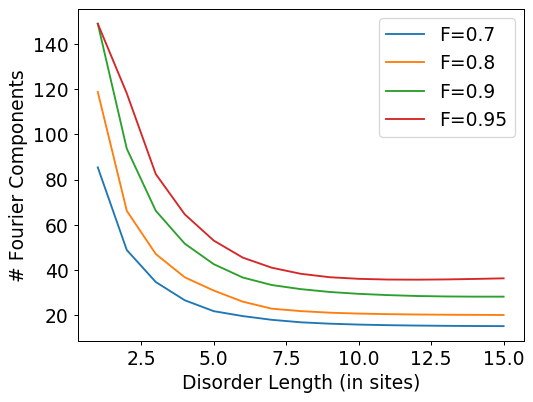}
         \caption{}
     \end{subfigure}
     \caption{{Number of disorder components to achieve different fidelities $F=\frac{\vec{V}_{dis}^{exact}\cdot \vec{V}_{dis}^{pred}}{\sqrt{(\vec{V}_{dis}^{exact}\cdot \vec{V}_{dis}^{exact})(\vec{V}_{dis}^{pred}\cdot \vec{V}_{dis}^{pred})}}$ for different wire and correlation lengths. The correlation lengths are written in terms of the number of sites, where lattice spacing  a = $10$nm.  The wires lengths were (a) $2\mu m$ (b) $3\mu m$ (c) $4\mu m$.  For a wire of length 3$\mu m$ and correlation length of 7 sites (70nm) 33 components are required to achieve an F=0.97.}}
     \label{fig:CompReq}
\end{figure*}

Training the GPU neural network required only {a single NVIDIA A100 for 1 hour, encompassing 30 epochs. This is not even close to the bottleneck and is effectively a negligible amount of resources. More importantly here we find that the neural network does not seem to obviously scale with the number of disorder components. It seems to be the scaling is more determined by the size of the input measurements (which we fix constant anyways) than the size of the output disorder components. There is likely some scaling in terms of the disorder components however it is clearly quite slow and certainly not exponential. This implies to us that the neural networks size is not a limiting factor at all. The results for this are shown in Fig. \ref{fig:ScalingNN}}. {While we show the scaling seems to work for our fixed number of measurements, as more components become necessary, higher K resolution may allow one to distinguish them more efficiently.} For larger systems, KWANT scales approximately linearly in the generation process, but a larger system may also demand more components. In cases where all or a significant number of disorder components are needed, it might be more efficient to employ a convolutional upsampling decoder \cite{percebois2023reconstructing} to produce the disorder function. This approach would likely outperform our current method, which is optimized for handling smaller numbers of components.

\newpage

\section{Additional results}
\label{S:Add}
This section presents additional plots of numerical results. We provide 10 component disorder potentials and their corresponding conductances for $V_{\text{Bias}}$ vs. $B_x$, conductances for $\mu$ vs. $B_x$ and $T_{IV}$ plots all for both the input and prediction Y realizations. We include the conductance plots because the machine learning model utilizes $\mu$ vs. $B_x$ as its main input during the training process. The $T_{IV}$ plots demonstrate the ML model's capability to predict $T_{IV}$ solely based on conductance measurements. Moreover, we provide the conductance and $T_{IV}$ results for a pristine sample, which readers may find useful for comparison.

\begin{figure*}[t]
     \centering
        \begin{subfigure}[b]{0.95\linewidth}
         \centering
         \includegraphics[width=\textwidth]{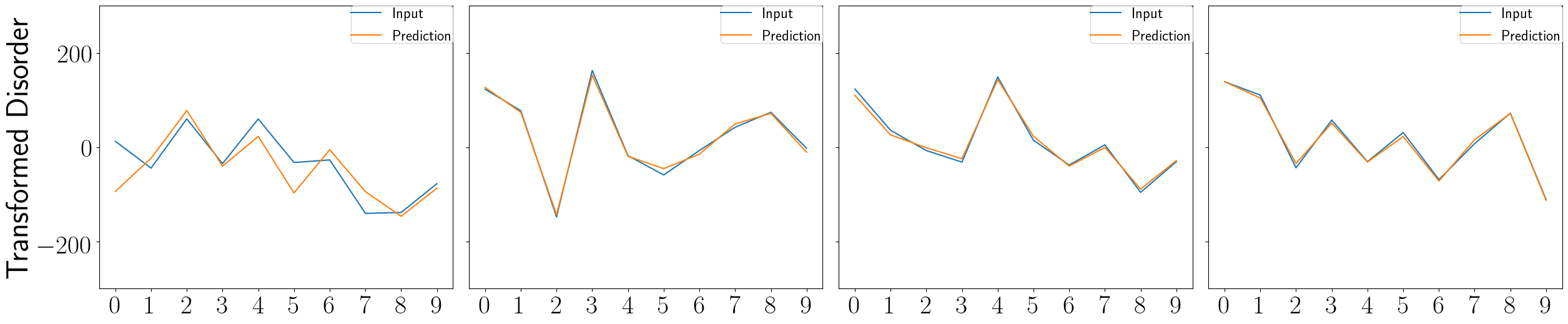}
         \caption{}
         \label{fig:5a}
     \end{subfigure}
         \begin{subfigure}[b]{0.95\linewidth}
         \centering
          \includegraphics[width=\textwidth]{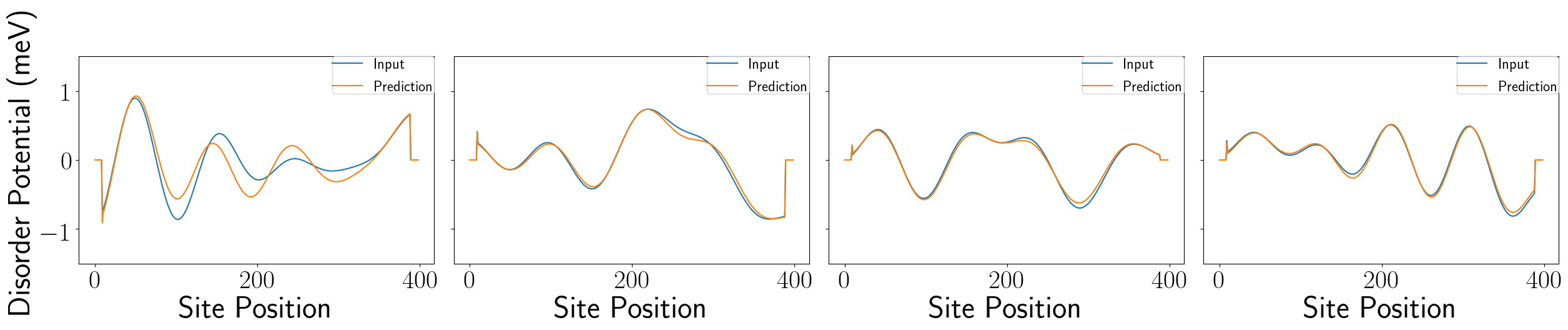}
         \caption{}
         \label{fig:5b}
     \end{subfigure}
     \begin{subfigure}[b]{0.48\linewidth}
         \centering
         \includegraphics[width=\textwidth]{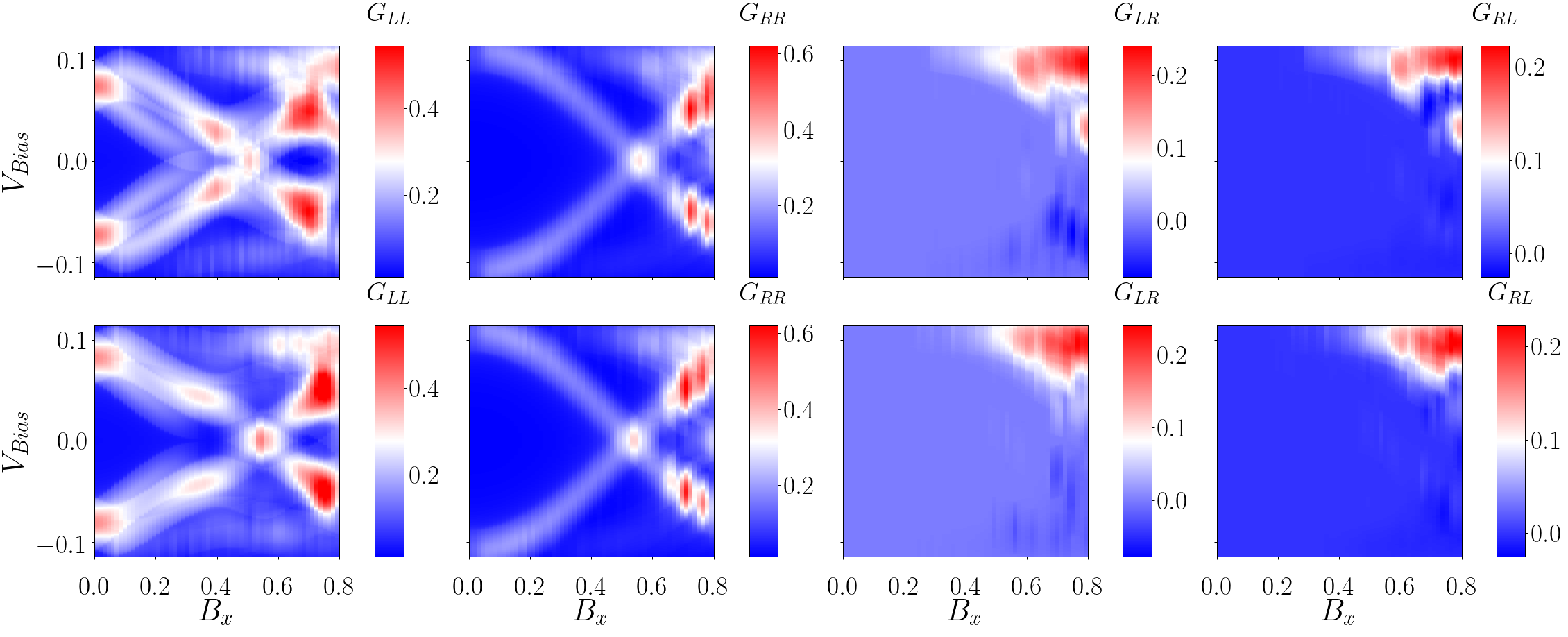}
         \caption{}
         \label{fig:5c}
     \end{subfigure}
          \begin{subfigure}[b]{0.48\linewidth}
         \centering
         \includegraphics[width=\textwidth]{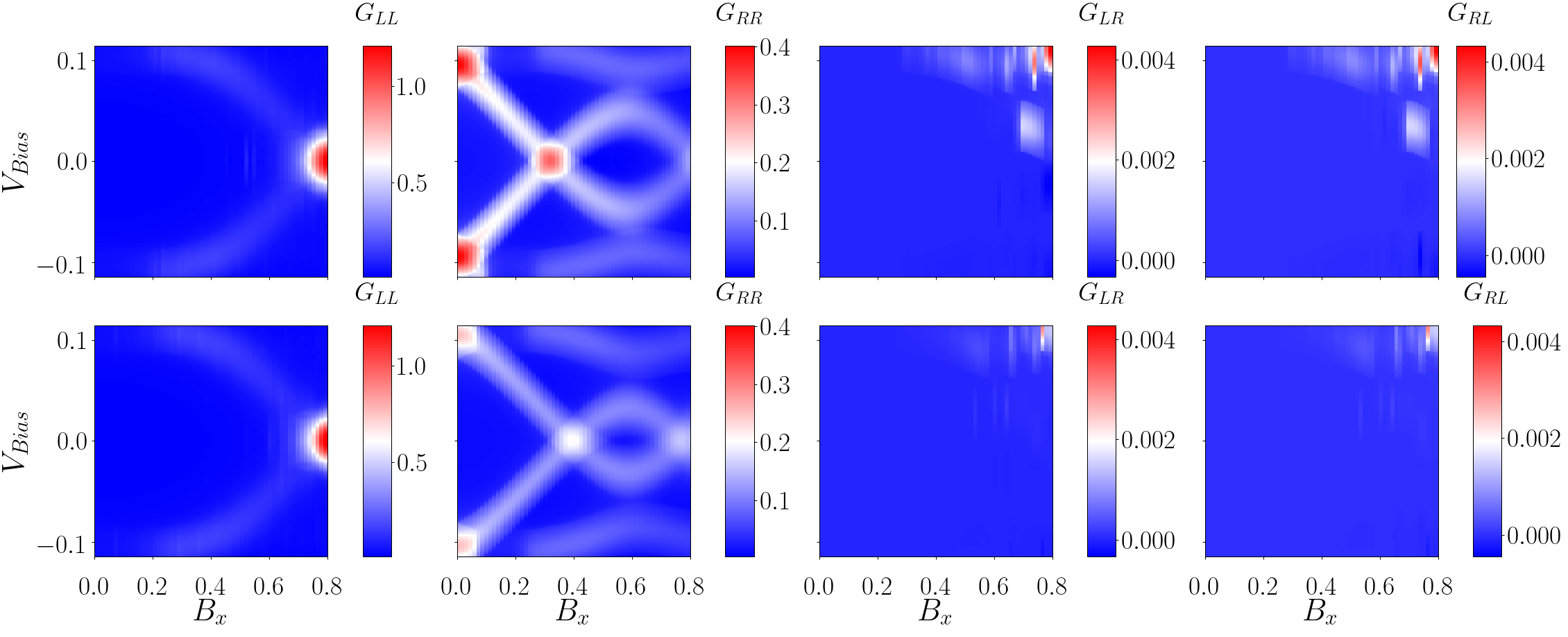}
         \caption{}
         \label{fig:5d}
     \end{subfigure}
          \begin{subfigure}[b]{0.48\linewidth}
         \centering
         \includegraphics[width=\textwidth]{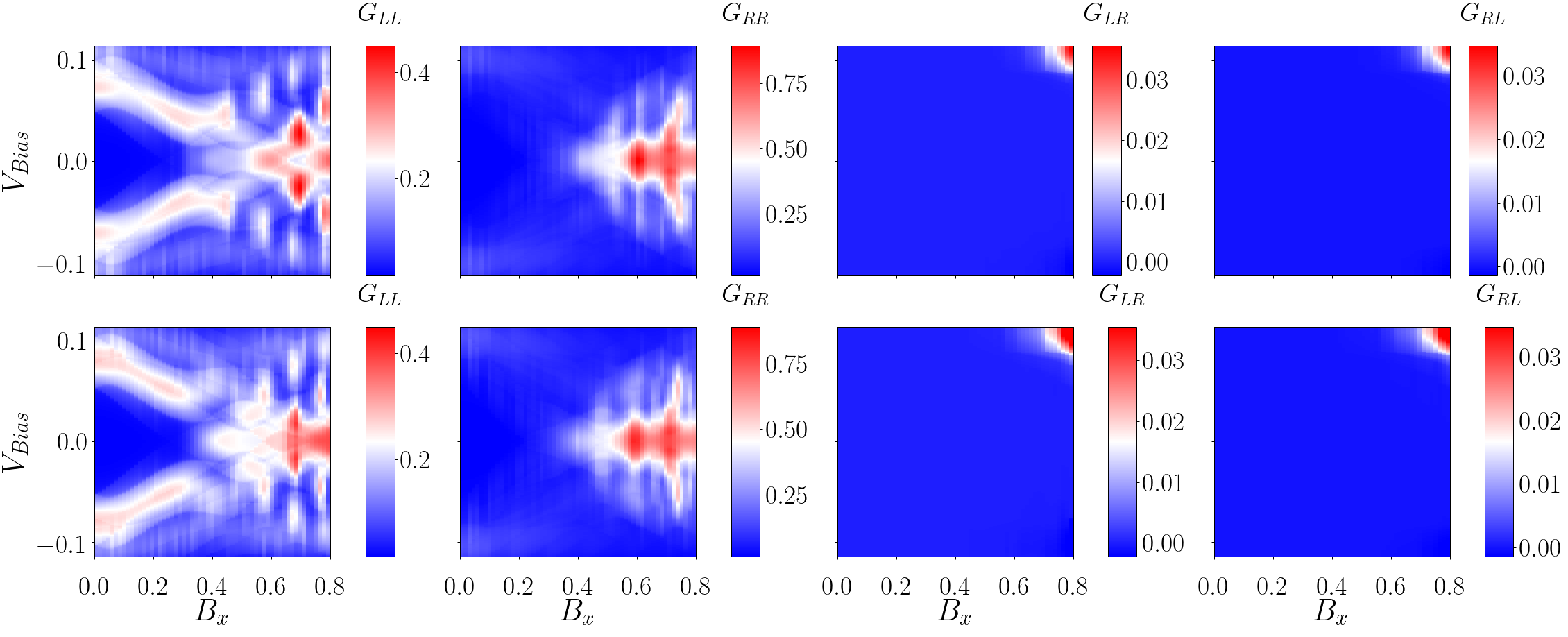}
         \caption{}
         \label{fig:5e}
     \end{subfigure}
          \begin{subfigure}[b]{0.48\linewidth}
         \centering
         \includegraphics[width=\textwidth]{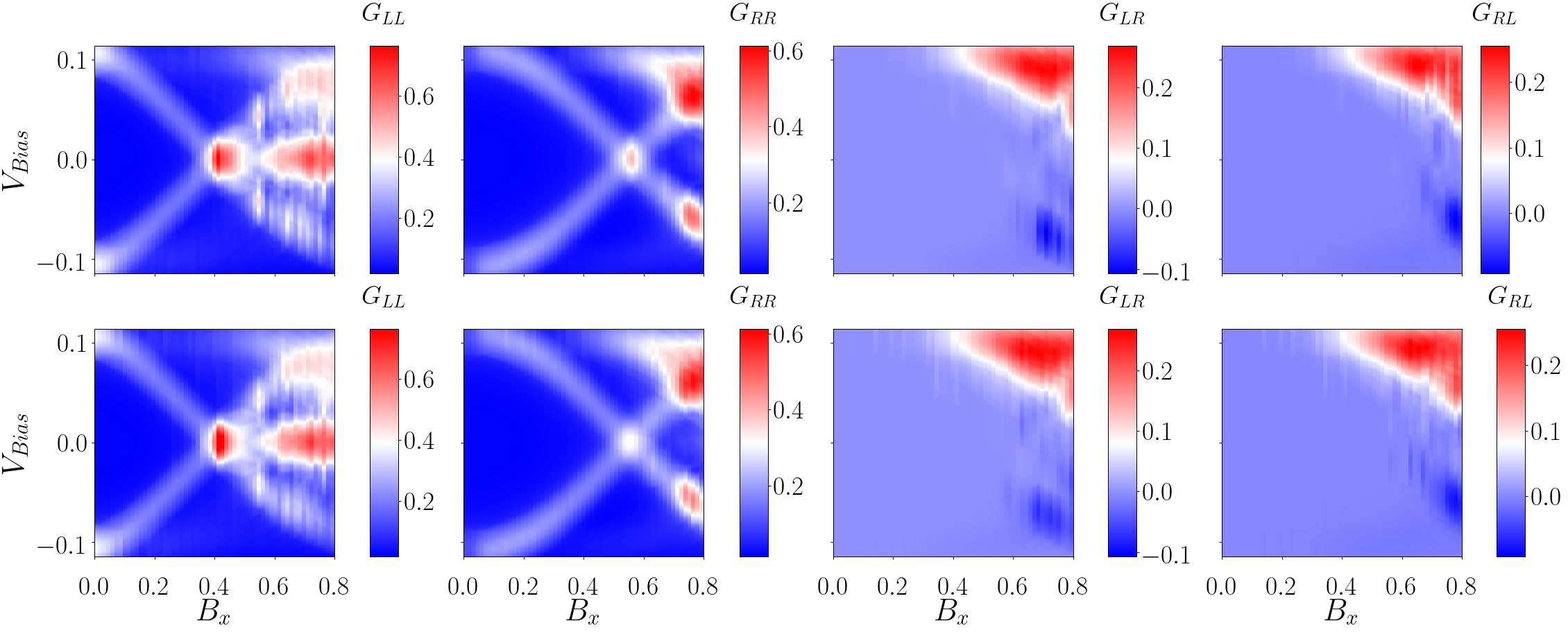}
         \caption{}
         \label{fig:5f}
     \end{subfigure}
     \caption{Representative samples of 10 disorder potential components neural network test input and predictions. (a) The discrete cosine transformed components of the disorder potential for 4 different samples and their respective predictions. (b) Disorder potentials in real space with deterministic modification due to boundary effects. {The longer correlation length (compared to $l=30\, nm$) of the disorder apparent in (b) is a result of limiting the number of fourier components.} In (c-f) the first row is $G_{LL}$, $G_{RR}$, $G_{LR}$ and $G_{RL}$ conductance measurements expected from a input disorder potential, and the second row is conductance measurements from the corresponding predicted disorder potential. Each of (c-f) are different disorder potential realizations, shown in (a). The conductance measurements are plotted for different values of $V_{\text{Bias}}$ (in meV) and $B_x$ (in T).   For this model {$\alpha=8 meV nm$} is fixed. The topological invariant corresponding to this profile can be seen in {color{red}Fig. S8}. The disorder amplitude and correlation length for each realization was $1.5$meV and 3 sites respectively. These are plotted for $\mu=0$. }
     \label{fig:dVB2}
\end{figure*}

\begin{figure*}[h]
     \centering
     \begin{subfigure}[b]{0.45\linewidth}
         \centering
         \includegraphics[width=\textwidth]{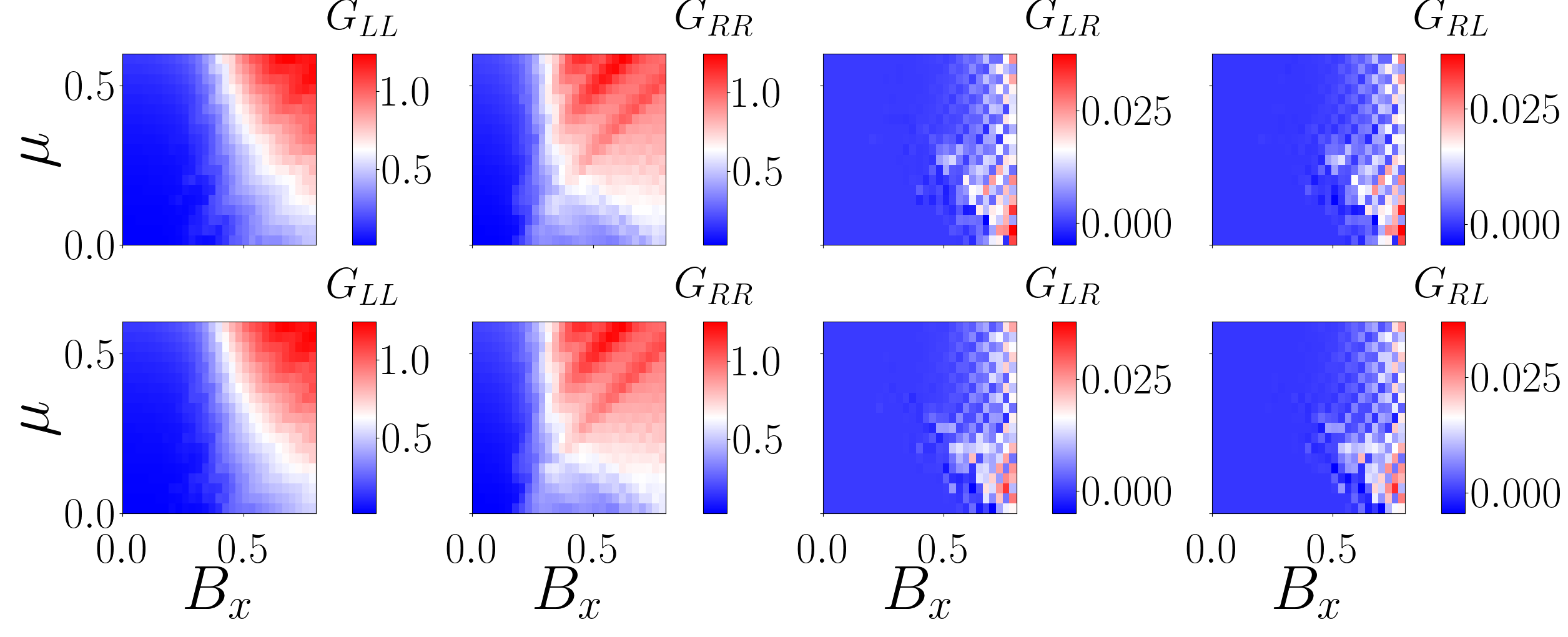}
         \caption{}
         \label{fig:S1c}
     \end{subfigure}
          \begin{subfigure}[b]{0.45\linewidth}
         \centering
         \includegraphics[width=\textwidth]{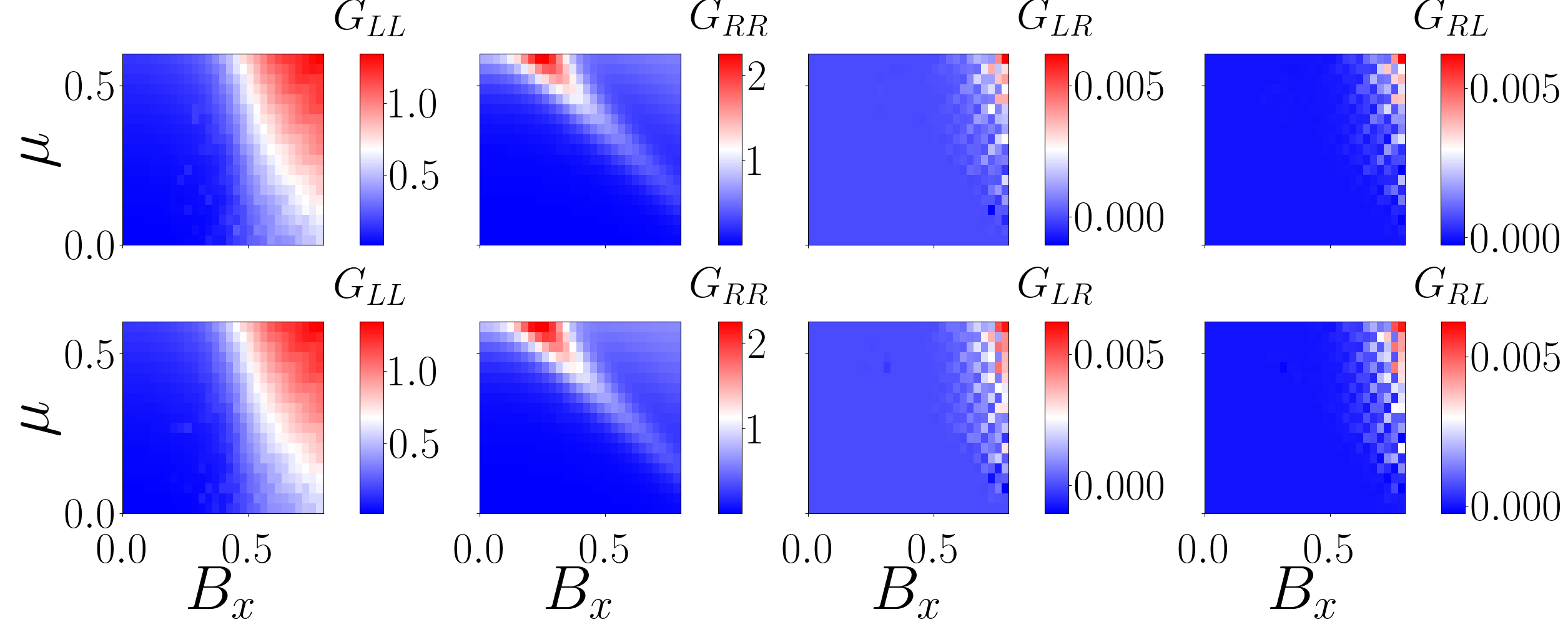}
         \caption{}
         \label{fig:S1d}
     \end{subfigure}
          \begin{subfigure}[b]{0.45\linewidth}
         \centering
         \includegraphics[width=\textwidth]{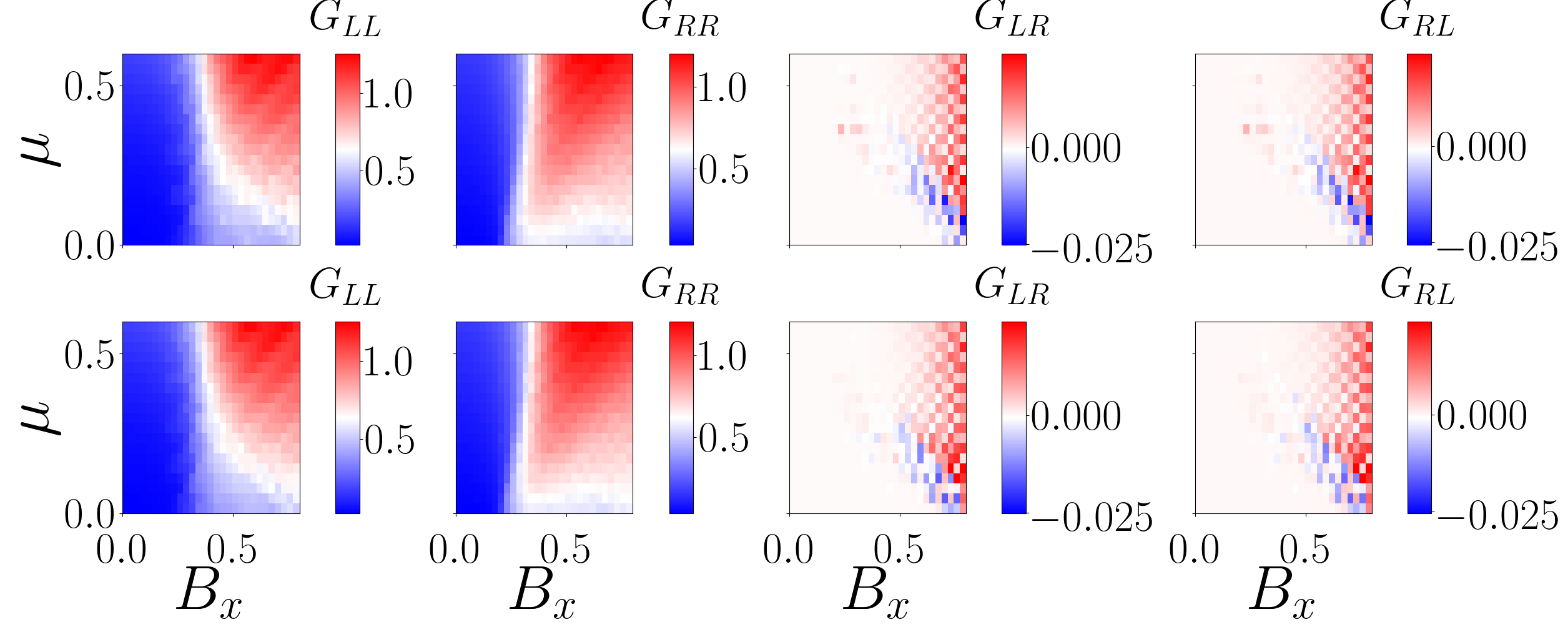}
         \caption{}
         \label{fig:S1e}
     \end{subfigure}
          \begin{subfigure}[b]{0.45\linewidth}
         \centering
         \includegraphics[width=\textwidth]{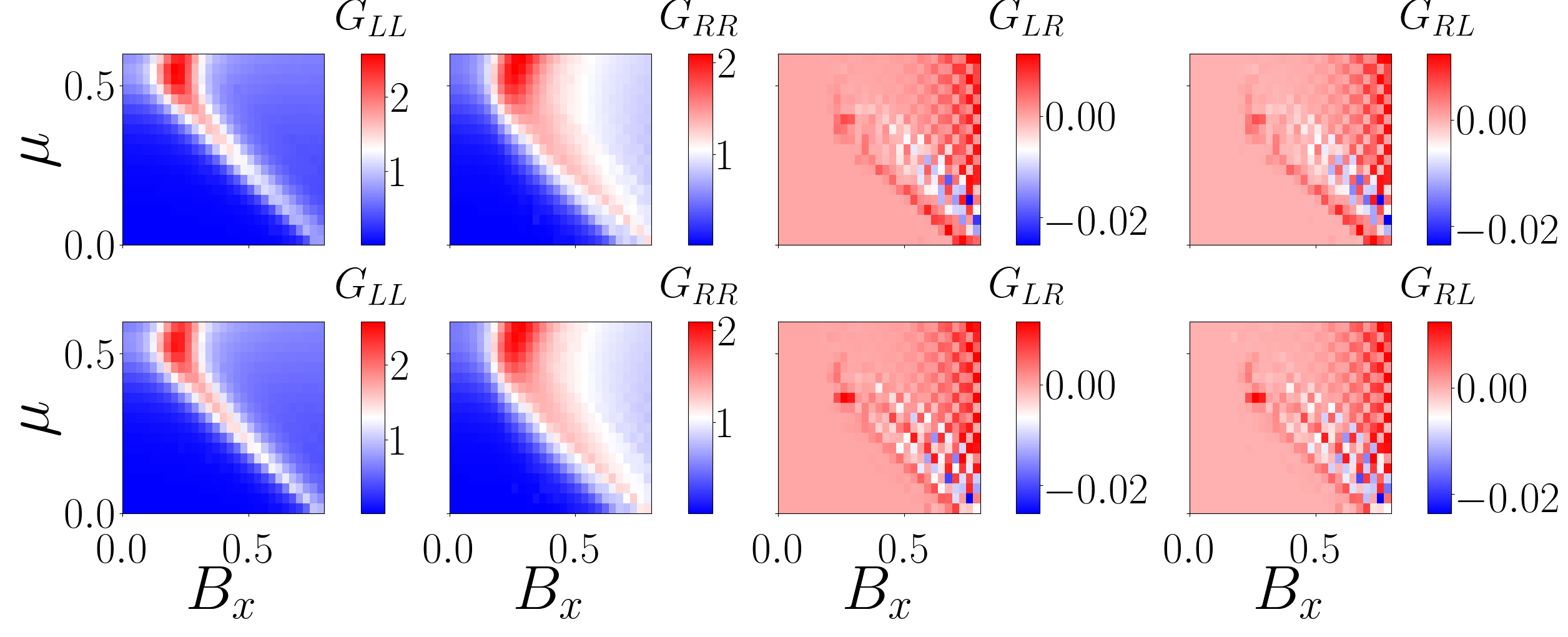}
         \caption{}
         \label{fig:S1f}
     \end{subfigure}
     \caption{{Representative samples of 5 disorder potential components} and $\alpha$ neural network test input and predictions. In (a-d) the first row is $G_{LL}$, $G_{RR}$, $G_{LR}$ and $G_{RL}$ conductance measurements expected from a input disorder potential, and the second row is conductance measurements from the corresponding predicted disorder potential. Each of (a-d) is a different Y realization, shown in Fig. \ref{fig:D1}. The conductance measurements are plotted for different values of $\mu$ (in meV) and $B_x$ (in T). The topological invariant corresponding to this profile can be seen in Fig. \ref{fig:VIT1}. The disorder amplitude and correlation length for each realization was 1.5meV and 3 sites respectively. These plots use $V_{\text{Bias}}=0$.}
     \label{fig:DC1}
\end{figure*}
\begin{figure*}[h]
     \centering
     \begin{subfigure}[b]{0.45\linewidth}
         \centering
         \includegraphics[width=\textwidth]{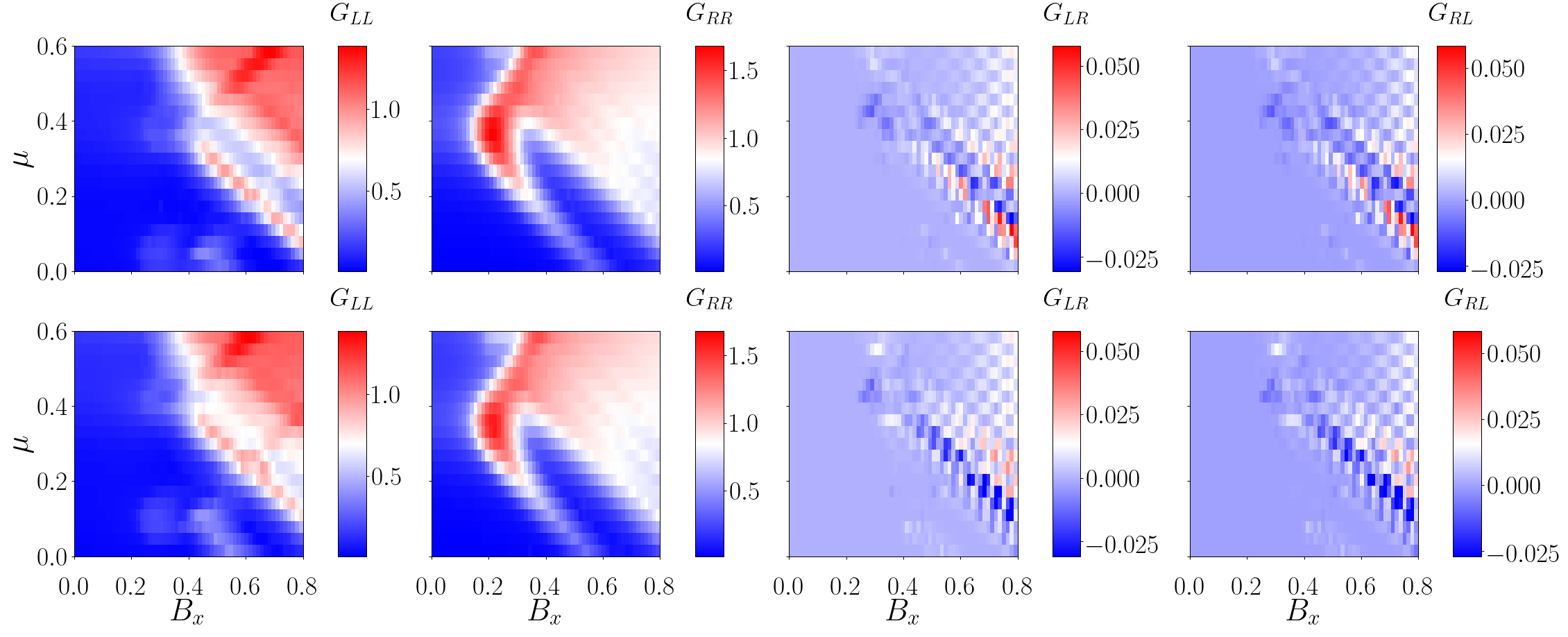}
         \caption{}
         \label{fig:S2c}
     \end{subfigure}
          \begin{subfigure}[b]{0.45\linewidth}
         \centering
         \includegraphics[width=\textwidth]{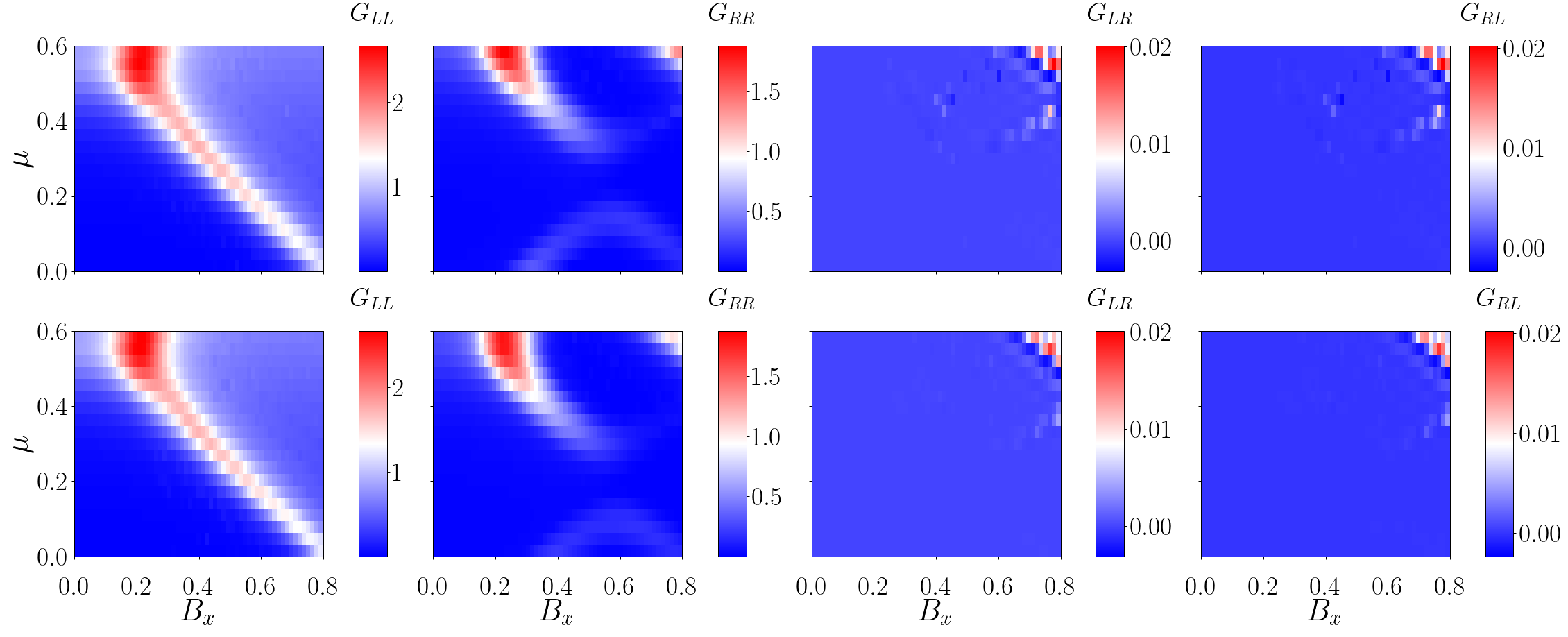}
         \caption{}
         \label{fig:S2d}
     \end{subfigure}
          \begin{subfigure}[b]{0.45\linewidth}
         \centering
         \includegraphics[width=\textwidth]{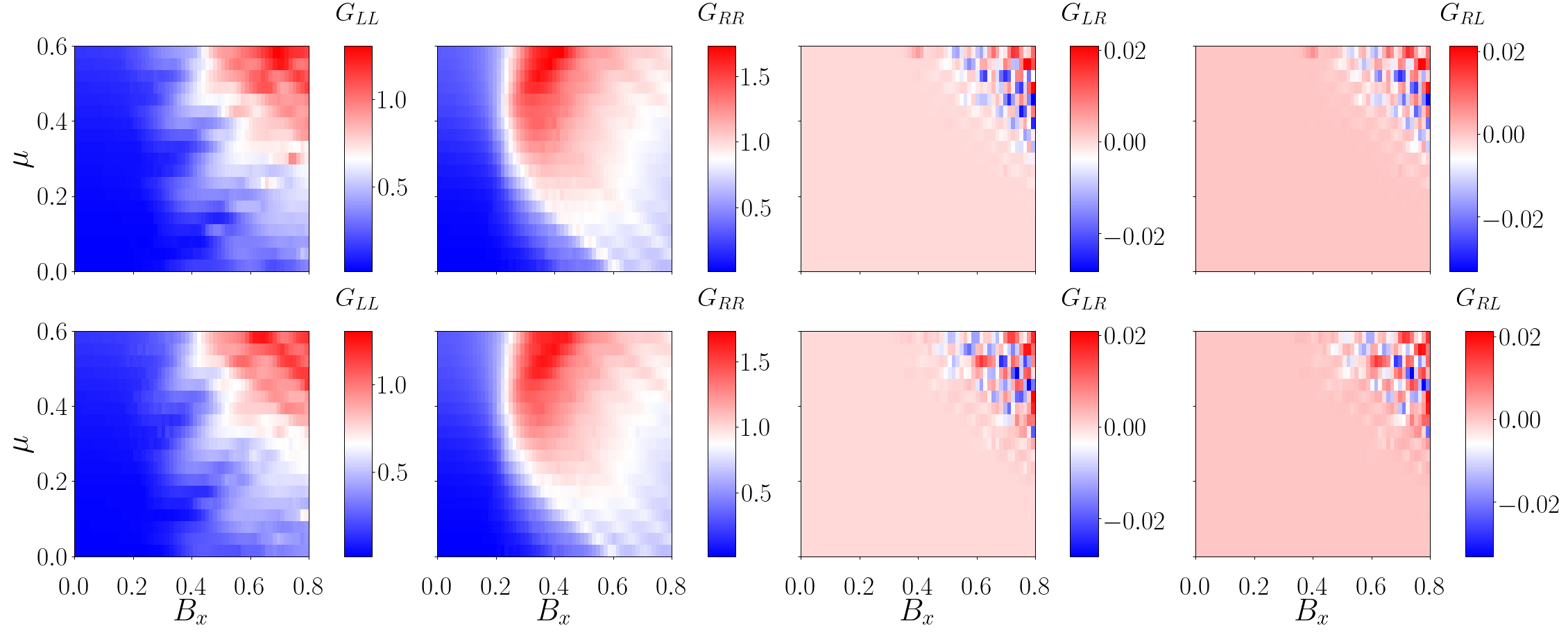}
         \caption{}
         \label{fig:S2e}
     \end{subfigure}
          \begin{subfigure}[b]{0.45\linewidth}
         \centering
         \includegraphics[width=\textwidth]{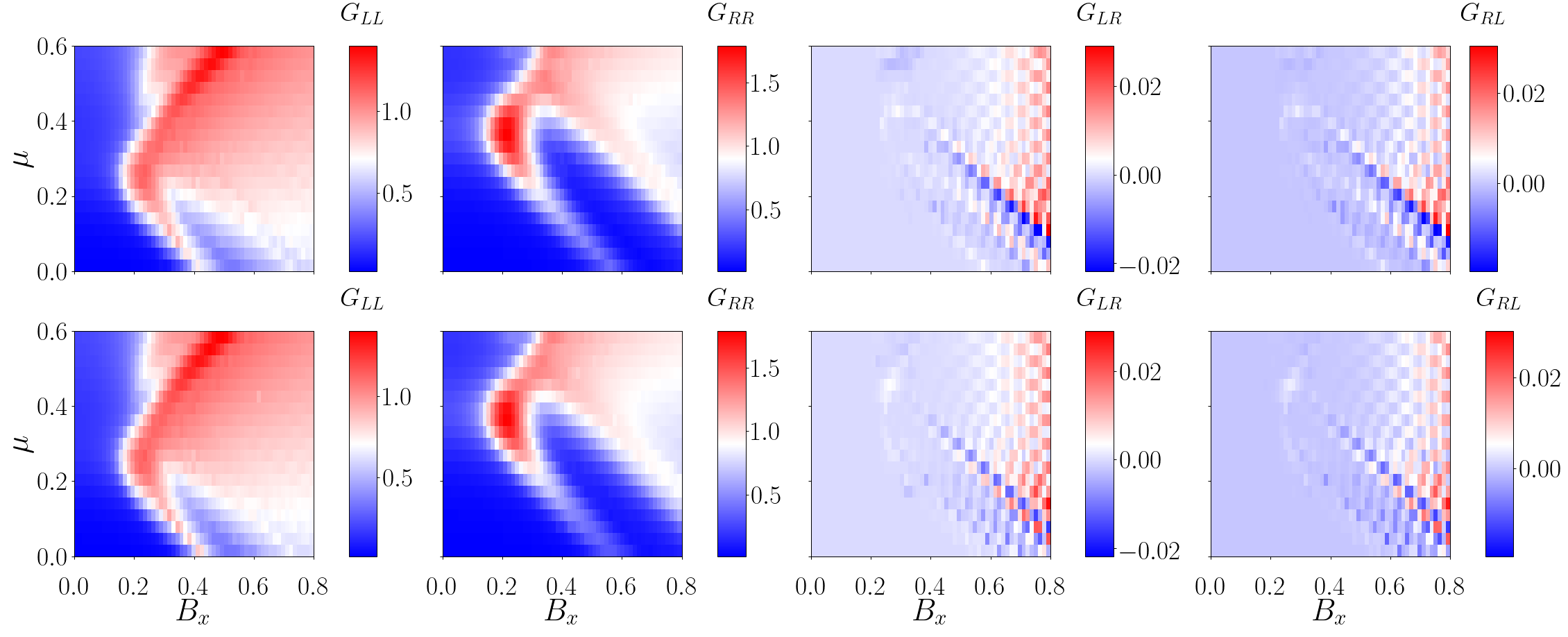}
         \caption{}
         \label{fig:S2f}
     \end{subfigure}
     \caption{Representative samples of 10 disorder potential components neural network test input and predictions. In (a-d) the first row is $G_{LL}$, $G_{RR}$, $G_{LR}$ and $G_{RL}$ conductance measurements expected from a input disorder potential, and the second row is conductance measurements from the corresponding predicted disorder potential. Each of (a-d) are different disorder potential realizations, shown in Fig. \ref{fig:dVB2}a. The conductance measurements are plotted for different values of $\mu$ (in meV) and $B_x$ (in T). For this model $\alpha=8 \text{meV nm}$ is fixed. The topological invariant corresponding to this profile can be seen in Fig. \ref{fig:VIT2}. The disorder amplitude and correlation length for each realization was 1.5meV and 3 sites respectively. These plots use $V_{\text{Bias}}=0$.}
     \label{fig:rDC2}
\end{figure*}
\begin{figure*}[h]
     \centering
     \begin{subfigure}[b]{0.9\linewidth}
         \centering
         \includegraphics[width=\textwidth]{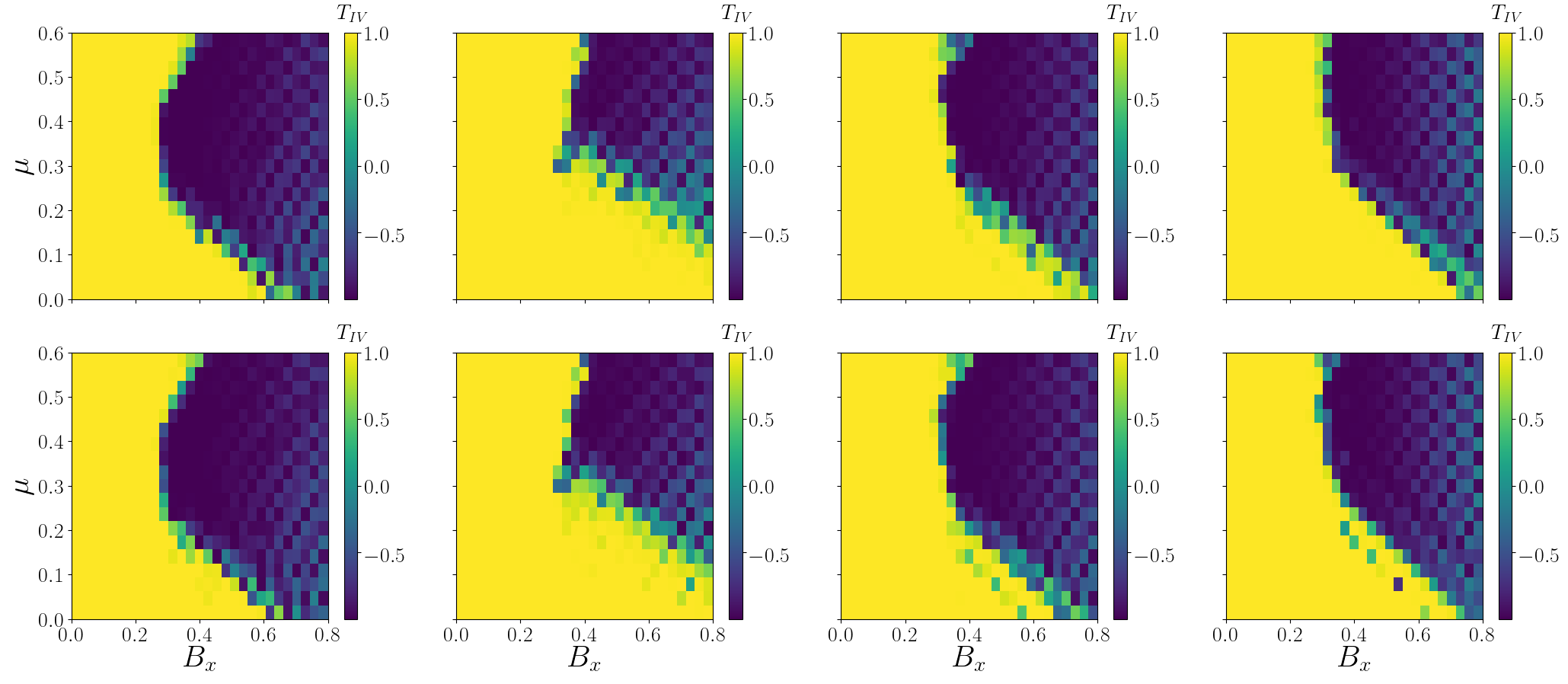}
         \caption{}
         \label{fig:9a}
     \end{subfigure}
     \caption{Representative samples of the topological invariant ($T_{IV}$) for 5 disorder potential components and $\alpha$ neural network test input and predictions. The first row is $T_{IV}$ resulting from a input disorder potential, and the second row is from the corresponding predicted disorder potential.  Each column are the different disorder potential realizations, shown in {Fig. \ref{fig:D1}}. The topological invariant is plotted for different values of $\mu$ (in meV) and $B_x$ (in T).}
     \label{fig:VIT1}
\end{figure*}

\begin{figure*}[h]
     \centering
     \begin{subfigure}[b]{0.9\linewidth}
         \centering
         \includegraphics[width=\textwidth]{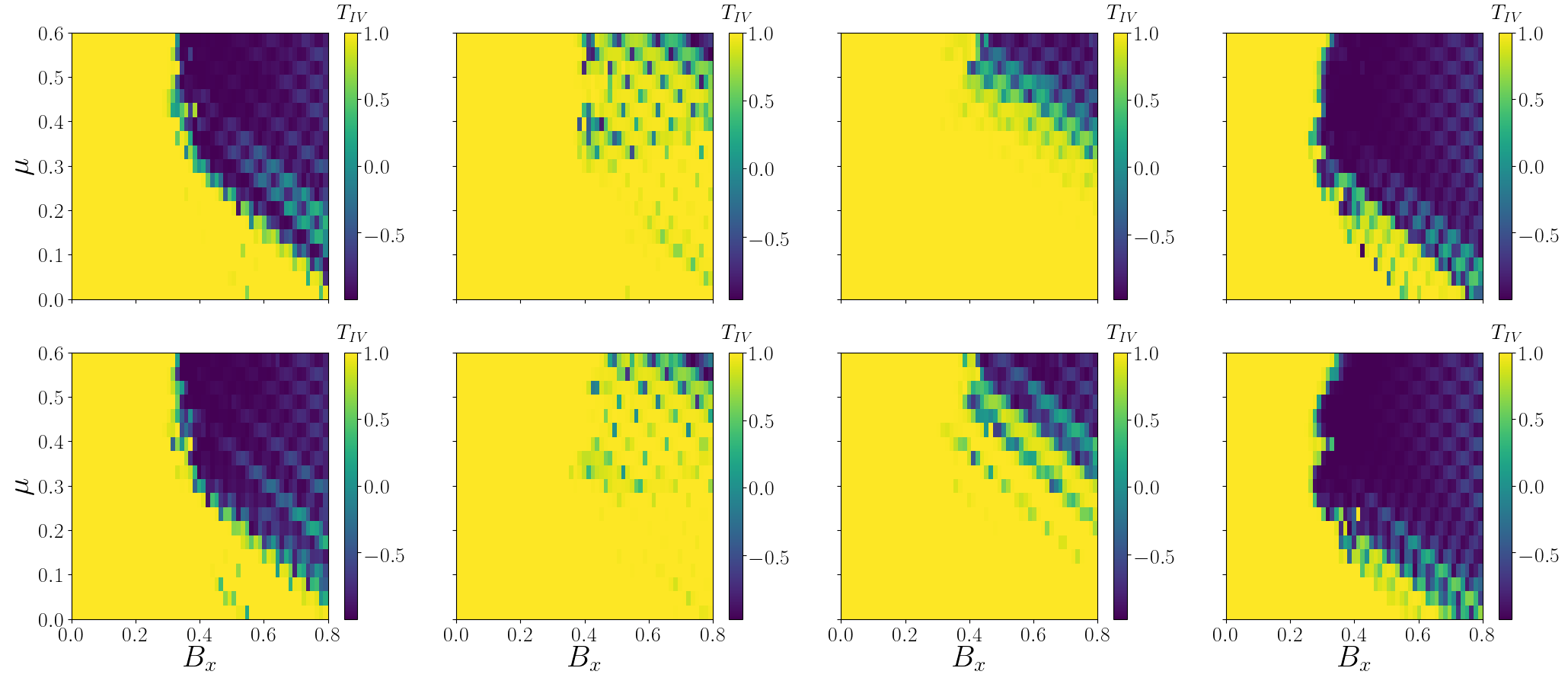}
         \caption{}
         \label{fig:S3a}
     \end{subfigure}
     \caption{Representative samples of the topological invariant ($T_{IV}$) for 10 disorder potential components neural network test input and predictions. The first row is $T_{IV}$ resulting from a input disorder potential, and the second row is from the corresponding predicted disorder potential. Each column are the different disorder potential realizations, shown in Fig. \ref{fig:dVB2}a. The dotted red line corresponds to the conductance profiles in Fig. \ref{fig:dVB2}b-e.  The topological invariant is plotted for different values of $\mu$ (in meV) and $B_x$ (in T).}
     \label{fig:VIT2}
\end{figure*}

\begin{figure*}[h]
     \centering
     \begin{subfigure}[b]{0.45\linewidth}
         \centering
         \includegraphics[width=\textwidth]{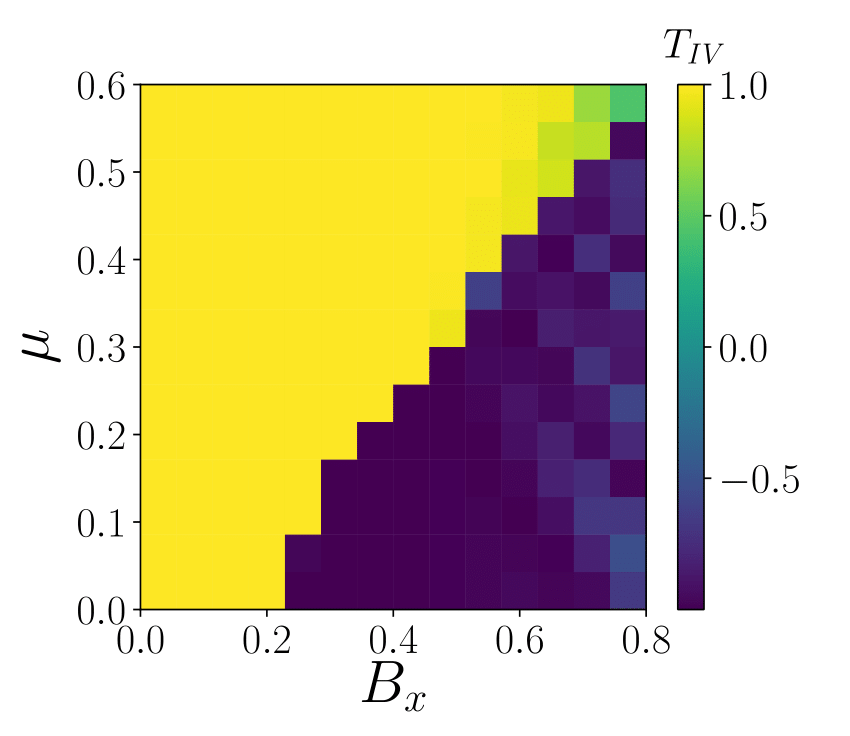}
         \caption{}
         \label{fig:S4a}
     \end{subfigure}
    \begin{subfigure}[b]{0.9\linewidth}
         \centering
         \includegraphics[width=\textwidth]{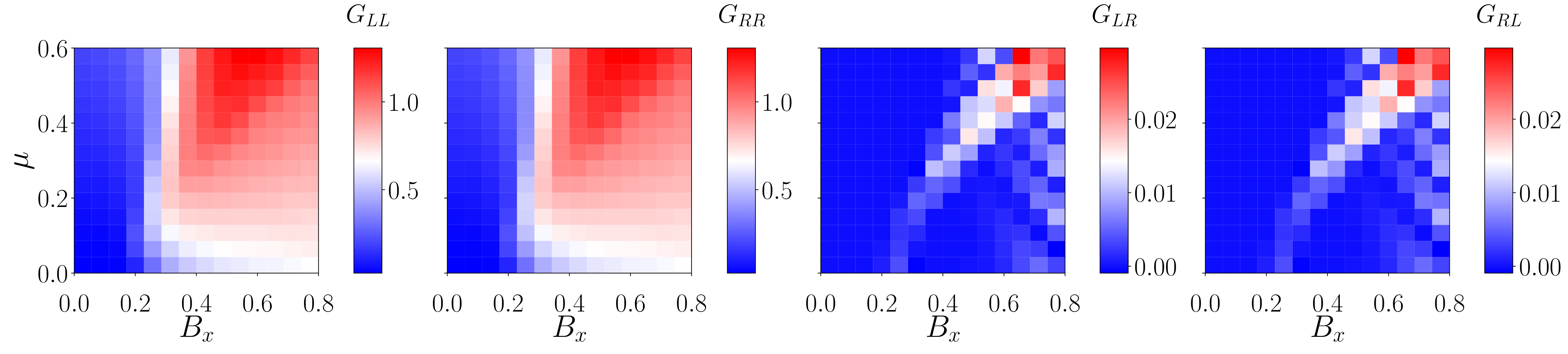}
         \caption{}
         \label{fig:S4b}
     \end{subfigure}
     \begin{subfigure}[b]{0.9\linewidth}
         \centering
         \includegraphics[width=\textwidth]{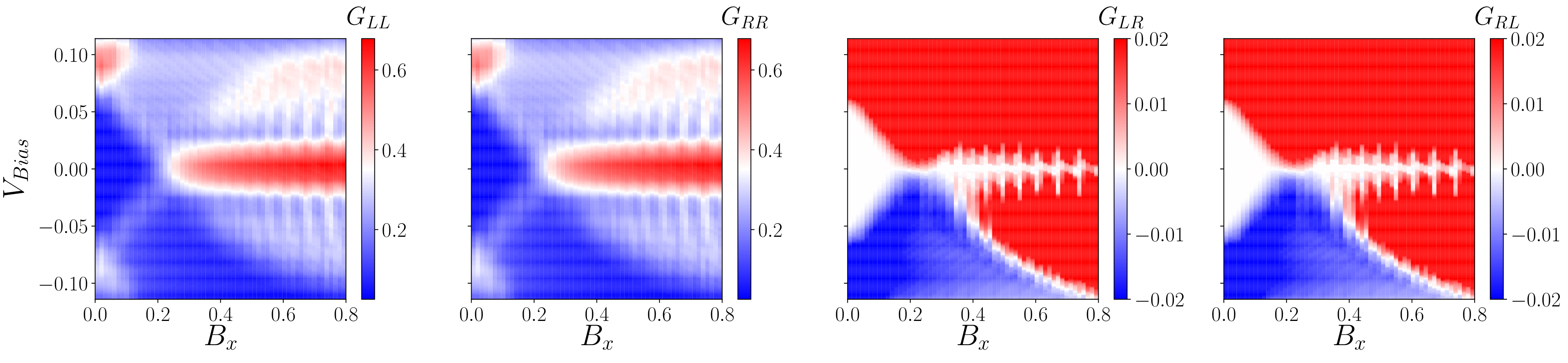}
         \caption{}
         \label{fig:S4c}
     \end{subfigure}
      \begin{subfigure}[b]{0.9\linewidth}
         \centering
         \includegraphics[width=\textwidth]{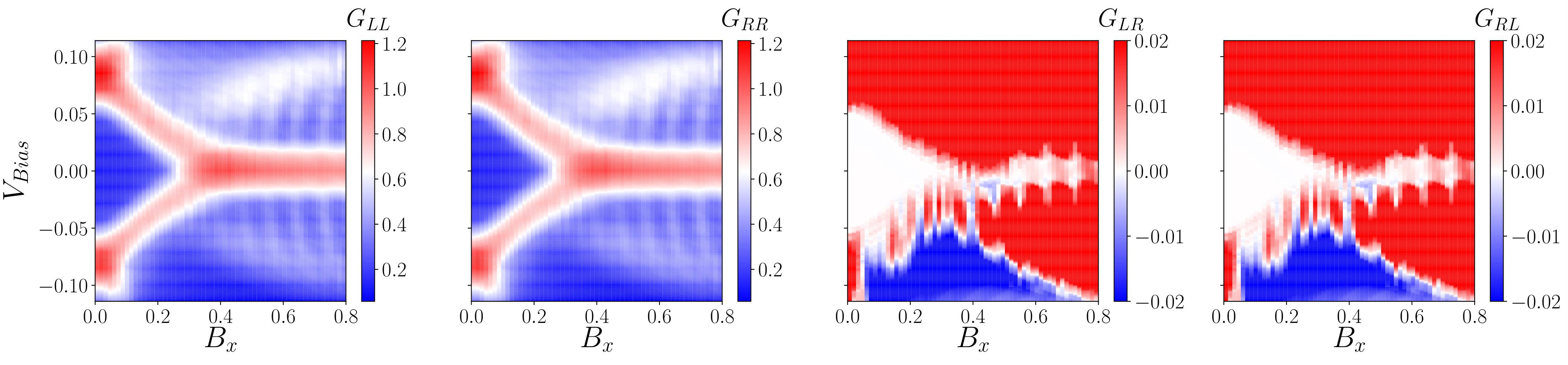}
         \caption{}
         \label{fig:S4d}
     \end{subfigure}
     \caption{Pristine sample ($V_{\text{dis}}(x)=0$) topological invariant ($T_{IV}$) and conductance measurements. (a) The topological invariant and (b) Conductance measurements are plotted for different values of $\mu$ (in meV) and $B_x$ (in T) at $V_{Bias}=0$. Conductance measurements at (c) $\mu=0$ and (d) $\mu=0.3$ meV are plotted for different values of $V_{\text{Bias}}$ (in meV) and $B_x$ (in T).  The plots held fixed {$\alpha=8 \text{meV nm}$}.}
     \label{fig:VIT}
\end{figure*}

\clearpage
\putbib
\end{bibunit}

\end{document}